\documentclass[]{iopart}
\usepackage{iopams, graphicx}
\usepackage{hyperref}  

\newcommand{\tfrac}[2]{\textstyle \frac{#1}{#2}}
\newcommand{\half}{\tfrac{1}{2}}
\newcommand{\third}{\tfrac{1}{3}}
\newcommand{\quarter}{\tfrac{1}{4}}
\newcommand{\Lie}{\mathcal{L}}
\newcommand{\Kzr}{{K^z{}_r}}
\newcommand{\tKzr}{{\tilde K^z{}_r}}
\newcommand{\tU}{\tilde U}
\newcommand{\tW}{\tilde W}

\begin{document}

\title[Constrained evolution in axisymmetry]
  {Constrained evolution in axisymmetry and the 
   gravitational collapse of prolate Brill waves}
\author{Oliver Rinne $^{1,2,3}$}
\address{$^1$ Department of Applied Mathematics and Theoretical Physics,
  Centre for Mathematical Sciences, Wilberforce Road, Cambridge CB3
  0WA, UK \\ $^2$ King's College, Cambridge CB2 1ST, UK \\
  $^3$ Theoretical Astrophysics 130-33, California Institute of Technology,
  1200 East California Boulevard, Pasadena, CA 91125, USA
}
\ead{O.Rinne@damtp.cam.ac.uk}

\begin{abstract}
  This paper is concerned with the Einstein equations in axisymmetric 
  vacuum spacetimes. 
  We consider numerical evolution schemes that solve the constraint
  equations as well as elliptic gauge conditions at each time step.
  We examine two such schemes that have been proposed in the literature and
  show that some of their elliptic equations are indefinite, thus
  potentially admitting nonunique solutions and causing numerical
  solvers based on classical relaxation methods to fail.
  A new scheme is then presented that does not suffer from these problems.
  We use our numerical implementation to study the gravitational
  collapse of Brill waves. 
  A highly prolate wave is shown to form a black hole rather than a 
  naked singularity.
\end{abstract}


%%%%%%%%%%%%%%%%%%%%%%%%%%%%%%%%%%%%%%%%%%%%%%%%%%%%%%%%%%%%%%%%%%%%%%%%%%%%%%%

\section{Introduction}
\label{s:Intro}

Driven by the need of gravitational wave data analysis for waveform 
templates, numerical relativity has focused in recent years on the 
modelling of astrophysical sources of gravitational waves such as the 
inspiral and coalescence of compact objects.
Such systems do not possess any symmetries and thus require a fully
$3+1$ dimensional numerical code.
The advantage of assuming a spacetime symmetry, on the other hand, is
that it allows for a dimensional reduction of the Einstein equations, 
which reduces the computational effort considerably so that greater
numerical accuracy can be obtained.
While spherical or planar symmetry yields the greatest reduction in
computational cost, the intermediate case of axisymmetry is more 
interesting in that it permits the study of gravitational waves.

In this article we focus on vacuum axisymmetric spacetimes and assume 
that the Killing vector is hypersurface orthogonal so that there is only one
gravitational degree of freedom.
The axisymmetric Einstein equations can be simplified considerably by
choosing suitable gauge conditions. 
Here we consider a combination of maximal slicing and quasi-isotropic
gauge \cite{Wilson1979,Bardeen1983,Evans1986}.
This gauge reduces the number of dependent variables to such an extent
that only one pair of evolution equations corresponding
to the one gravitational degree of freedom needs to be kept. 
All the other variables can be solved for using the constraint equations and
gauge conditions. This \emph{fully constrained} approach was taken 
in \cite{Choptuik2003}. \emph{Partially constrained} schemes 
(e.g., \cite{Abrahams1993,Garfinkle2001})
substitute some of the constraint equations with evolution equations; 
this is possible because the Einstein equations are overdetermined.

Such (fully or partially) constrained schemes have proven very robust
in simulations of strong gravity phenomena. Examples include the
collapse of vacuum axisymmetric gravitational waves, so-called Brill
waves \cite{Brill1959}, in \cite{Garfinkle2001}. Critical phenomena in
this system were found in \cite{Abrahams1993,Abrahams1994}. 
Critical phenomena in the collapse of massless scalar 
fields \cite{Choptuik2003b} and complex scalar fields with angular 
momentum \cite{Choptuik2004} were also studied, as was the collapse 
of collisionless matter \cite{Shapiro1991}.

Nevertheless, constrained evolution schemes have been plagued with
problems. The authors of \cite{Choptuik2003} reported that their
multigrid elliptic solver failed occasionally for the Hamiltonian
constraint equation in the strong-field regime. This problem can be
circumvented by using instead the evolution equation for the conformal
factor. However, it was found that this was ``not sufficient to ensure
convergence in certain Brill-wave dominated spacetimes''.
Similar difficulties were encountered in \cite{BarnesPhD,RinnePhD}. 
The purpose of this article is to determine the cause of these
problems and to develop an improved constrained evolution scheme.

The suspect elliptic equations belong to a class of (nonlinear)
Helmholtz-like equations, which are discussed quite generally in
section \ref{s:Ell}. We point out that if these Helmholtz equations
are \emph{indefinite} (loosely speaking, they have the ``wrong sign'')
then their solutions, should they exist, are potentially nonunique. 
The same criterion is found to be related to the convergence of 
numerical solvers based on classical relaxation methods.

In section \ref{s:Formulation}, we review the partially constrained
scheme of \cite{Garfinkle2001} and the fully constrained scheme 
of \cite{Choptuik2003}.
We show that some of the elliptic equations in these formulations are
indefinite. This leads us to the construction of a modified fully
constrained scheme that does not suffer from this problem.
The arguments involved turn out to be closely related to questions 
of (non)uniqueness in conformal approaches to the initial data 
problem in standard $3+1$ numerical relativity \cite{Pfeiffer2005,Walsh2007}.

A numerical implementation of the new fully constrained scheme is
described in section \ref{s:NumMethod}. In section \ref{s:NumResults},
we apply it to a study of Brill wave gravitational collapse.
After performing a convergence test and comparing results for a strong 
wave with ``spherical'' initial data, we focus on a highly prolate 
configuration--one of the initial data sets examined 
in \cite{Abrahams1992}. 
By considering sufficiently prolate configurations, the authors were able 
to construct initial data without an apparent horizon but apparently with 
arbitrarily large curvature. They conjectured that such initial data
would evolve to form a naked singularity rather than a black hole.
This would constitute a violation of weak cosmic censorship.
A numerical evolution of one of these prolate initial data sets 
was carried out in \cite{Garfinkle2001}. 
Due to a lack of resolution on their compactified spatial grid, 
the authors could not evolve the wave for a sufficiently long time.
The trends in certain quantities suggested however that an apparent 
horizon would eventually form.
Using our new constrained evolution scheme, we are able to evolve the 
same initial data for much longer and we confirm that an apparent 
horizon does form.

We conclude and discuss some open questions in section \ref{s:Concl}.

%%%%%%%%%%%%%%%%%%%%%%%%%%%%%%%%%%%%%%%%%%%%%%%%%%%%%%%%%%%%%%%%%%%%%%%%%%%%%%%

\section{General remarks on Helmholtz-like elliptic equations}
\label{s:Ell}

Let us first consider the \emph{Helmholtz equation}
\begin{equation}
  \label{e:LinearHelmholtz}
  \Delta u + c u = f,
\end{equation}
where $u \in \mathbb{R}^n$, $\Delta$ is the flat-space Laplacian, 
and $c$ and $f$ are smooth functions. 
We impose the boundary condition $u \rightarrow 0$
at spatial infinity. More generally, a boundary condition 
$u \rightarrow a = \mathrm{const}$ can always be transformed to this
case by considering the function $u - a$.
It follows from standard elliptic theory (see e.g.~\cite{Gilbarg1977})
that \eref{e:LinearHelmholtz} has a unique solution if $c \leqslant 0$ 
everywhere. If $c > 0$ then multiple solutions may exist or there may
not be any solution at all. For $c > 0$ the elliptic equation
is said to be \emph{indefinite}.

Next we consider the quasilinear equation
\begin{equation}
  \label{e:NonlinearHelmholtz}
  \Delta u + F(u) = f,
\end{equation}
where $F(u)$ is a smooth (not necessarily linear) function.
Proving existence and uniqueness of solutions to this equation is
nontrivial. However a necessary condition for uniqueness can easily be
obtained. Suppose $u_0$ is a given solution and consider a small
perturbation of it, $u = u_0 + \delta u$. Approximating $F(u) \approx
F(u_0) + F'(u_0) \delta u$ where $F'(u) \equiv \rmd F(u) / \rmd u$,
we find that for $u$ to be a solution of \eref{e:NonlinearHelmholtz}, 
$\delta u$ must satisfy
\begin{equation}
  \label{e:LinearizedNonlinearHelmholtz}
  \Delta \delta u + F'(u_0) \delta u = 0.
\end{equation}
This is of the form \eref{e:LinearHelmholtz} with $c = F'(u_0)$ and $f
= 0$. If $F'(u_0) \leqslant 0$ then there is only the trivial solution
$\delta u \equiv 0$ and we call the original problem 
\eref{e:NonlinearHelmholtz} \emph{linearization stable}. 
If on the other hand $F'(u_0) > 0$ then multiple solutions of the 
linearized problem \eref{e:LinearizedNonlinearHelmholtz} 
and hence also of the nonlinear problem \eref{e:NonlinearHelmholtz} may exist.

As an example relevant to the formulations of the Einstein equations
discussed in this article, we take 
\begin{equation}
  \label{e:NonlinearHelmholtzExample}
  F(u) = c u^p
\end{equation}
with $p \in \mathbb{R}$ and $c$ a smooth function.
Then \eref{e:NonlinearHelmholtz} is linearization stable provided that
$pc \leqslant 0$. We say that in this case the equation 
has the ``right sign''.

Because of the above considerations on the uniqueness of solutions, 
it is clearly desirable to have an equation with the ``right sign'' if
a numerical solution is attempted. There is however also a more
practical reason. Consider again the linear Helmholtz equation
\eref{e:LinearHelmholtz} in, say, $n = 2$ dimensions.
Suppose we cover the domain with a uniform Cartesian grid with spacing $h$,
denoting the value of $u$ at the grid point with indices $(i,j)$ by $u_{ij}$.
A discretization of \eref{e:LinearHelmholtz} using second-order accurate
centred finite differences yields
\begin{equation}
  \label{e:ModelFD}
  u_{i+1,j} + u_{i-1,j} + u_{i,j+1} + u_{i,j-1} 
    - (4 - c h^2) u_{ij} = h^2 f_{ij}.
\end{equation}
We formally write this system of linear equations as
\begin{equation}
  \label{e:LinearSystem}
  A \mathbf{u} = h^2 \mathbf{f}, \quad \mathbf u \equiv \{ u_{ij} \}.
\end{equation}
Large systems \eref{e:LinearSystem} are commonly solved using
relaxation methods, which obtain a series of successively improved
numerical approximations. For example, a step of the \emph{Gauss-Seidel} 
method consists in sweeping through the grid (typically in
lexicographical or in red-black order), 
at each grid point $(i,j)$ 
solving the equation for $u_{ij}$ and replacing its value,
\begin{equation}
  u_{ij} \leftarrow 
  (u_{i+1,j} + u_{i-1,j} + u_{i,j+1} + u_{i,j-1} - h^2 f_{ij})
  /(4 - c h^2).
\end{equation}
The relaxation converges provided the matrix $A$ in \eref{e:LinearSystem}
is \emph{strictly diagonally dominant}, i.e., in each row of the
matrix the absolute value of the diagonal term is greater than the sum
of the absolute values of the off-diagonal terms (see e.g.~\cite{Stoer2002}).
In our example \eref{e:ModelFD}, the diagonal term is $|4 - ch^2|$ and
the off-diagonal terms add up to $4$, so that the condition for
convergence is $c < 0$. (The other possibilty, $c$ large and
positive, is not feasible because $h \rightarrow 0$ in the continuum limit.)
In practice, if $c$ is positive but sufficiently small then the
relaxation will still converge but as $c$ is increased 
convergence will begin to stall and ultimately the relaxation will diverge.
Similar convergence criteria hold for other relaxation schemes such as
the Jacobi or SOR methods.
In particular, the multigrid method \cite{Brandt1977} is based on 
these relaxation schemes and will not converge if the underlying
relaxation does not.
These warnings do not apply to certain versions of the conjugate 
gradient method \cite{Stoer2002} or other Krylov subspace iterations
which ideally only require the matrix $A$ to be invertible. 
For a combination of such methods with multigrid see e.g.~\cite{Elman2001}.

%%%%%%%%%%%%%%%%%%%%%%%%%%%%%%%%%%%%%%%%%%%%%%%%%%%%%%%%%%%%%%%%%%%%%%%%%%%%%%%

\section{Formulations of the axisymmetric Einstein equations}
\label{s:Formulation}

We focus on axisymmetric vacuum spacetimes. 
Axisymmetry means that there is an everywhere spacelike Killing vector 
field $\xi$ with closed orbits. 
Here we restrict ourselves to the case where the Killing vector is
hypersurface orthogonal.
We choose cylindrical polar coordinates $t, z, r, \phi$ such that
$\xi = \partial/\partial \phi$. 
In the following, indices $a,b,\ldots$ range over $t, r, z, \phi$, indices 
$i,j,\ldots$ over $r, z, \phi$, and indices $A,B,\ldots$ over $r, z$.
The line element is written in the form
\begin{equation}
  \rmd s^2 = -\alpha^2 \rmd t^2 + \psi^4 \{ \rme^{2rS} 
    [ (\rmd r - \beta^r \rmd t)^2 + (\rmd z - \beta^z \rmd t)^2 ]
    + r^2 \rmd \phi^2 \}.
\end{equation}
Here $\alpha$ and $\beta^A$ are the usual ADM lapse function and shift vector.

We have imposed as a gauge condition that the 2-metric on the $t =
\mathrm{const}, \phi = \mathrm{const}$ hypersurfaces be conformally
flat in our coordinates (quasi-isotropic gauge):
the spatial metric obeys $\gamma_{rz} = 0$ and $\gamma_{rr} = \gamma_{zz}$.
This condition must be preserved by the evolution equation for the
spatial metric,
\begin{equation}
  \partial_t \gamma_{ij} = - 2 \alpha K_{ij} + \Lie_\beta \gamma_{ij},
\end{equation}
where $K_{ij}$ is the extrinsic curvature of the $t = \mathrm{const}$
surfaces and $\Lie$ denotes the Lie derivative. We deduce that
\begin{eqnarray}
  \label{e:FirstOrderShift}
  \beta_+ \equiv \beta^z{}_{,r} + \beta^r{}_{,z} = 2 \alpha \Kzr,\nonumber\\
  \beta_- \equiv \beta^r{}_{,r} - \beta^z{}_{,z} = -\alpha U,
\end{eqnarray}
where $U \equiv K^z{}_z - K^r{}_r$.

Maximal slicing $K \equiv K_i{}^i = 0$ is imposed, so that the extrinsic
curvature has three degrees of freedom, which are taken to be
$\Kzr$, $U$, and $W \equiv (K^r{}_r - K^\phi{}_\phi)/r$
(this particular combination is motivated by regularity on the axis of
symmetry \cite{Garfinkle2001}).
The evolution equation for the extrinsic curvature is given by
\begin{equation}
  \partial_t K^i{}_j = - D^i D_j \alpha + \alpha R^i{}_j + \Lie_\beta K^i{}_j, 
\end{equation}
where $D$ is the covariant derivative compatible with the spatial metric
$\gamma_{ij}$ and $R_{ij}$ is its Ricci tensor.
Preservation of the maximal slicing condition implies $D_i D^i \alpha
= \alpha R = \alpha K^i{}_j K^j{}_i$ (using the Hamiltonian constraint
for the second equality) or
\begin{eqnarray}
  \label{e:GDSlicingCondition}
  \fl \alpha_{,rr} + \alpha_{,zz} + (2 P_r + r^{-1}) \alpha_{,r} + 2 P_z \alpha_{,z}
    \nonumber\\- 2\alpha \psi^4 \rme^{2rS} \left[ 
      \third (U + \half r W)^2 + \quarter (r W)^2 + (\Kzr)^2
    \right] = 0.
\end{eqnarray}
Here and in the following we use the notation 
\begin{equation}
  P_A \equiv \psi^{-1} \psi_{,A} , \qquad
  R_A \equiv (rS)_{,A} , \qquad
  A_A \equiv \alpha^{-1} \alpha_{,A} .
\end{equation}

There are many different ways of constructing an evolution scheme for
the axisymmetric Einstein equations in the above gauge, depending on
the number of constraint equations being solved. We review two schemes
that have been used for numerical simulations and show that some of their
elliptic equations are indefinite as discussed in the section \ref{s:Ell}. 
Finally we propose a new scheme that does not suffer from this problem.

\subsection{A partially constrained scheme: 
  Garfinkle and Duncan \cite{Garfinkle2001}}

Garfinkle and Duncan \cite{Garfinkle2001} choose to solve only the
Hamiltonian constraint equation $R - K^i{}_j K^j{}_i = 0$
(note that $K = 0$), which takes the form
\begin{eqnarray}
  \label{e:GDHamiltonianConstraint}
  \fl \psi_{,rr} + \psi_{,zz} + r^{-1} \psi_{,r} + \quarter [ (rS)_{,rr} + 
    (rS)_{,zz} ] \nonumber\\ + \quarter \psi^5 \rme^{2rS} \left[ 
    \third (U + \half r W)^2 + \quarter (r W)^2 + (\Kzr)^2
    \right] = 0.
\end{eqnarray}
This equation is of the type \eref{e:NonlinearHelmholtz}, 
\eref{e:NonlinearHelmholtzExample} with $p = 5$
and $c \geqslant 0$ (note the second square bracket in 
\eref{e:GDHamiltonianConstraint} is non-negative).
Hence it has the ``wrong sign'' and suffers from potential nonuniqueness 
of solutions as well as difficulties in solving it numerically using 
relaxation methods (section \ref{s:Ell}).
The latter is not a concern in \cite{Garfinkle2001} though because 
the authors use a conjugate gradient method.

The momentum constraints $D^j K_{ij} = 0$ are not solved but only 
monitored during the evolution. 
Written out explicitly they are
\begin{eqnarray}
  \label{e:GDMomentumConstraints}
  \fl 2 \Kzr_{,z} - \tfrac{2}{3} U_{,r} + \tfrac{2}{3} r W_{,r}
    + (12 P_z + 4 R_z) \Kzr - (4 P_r + 2 R_r) U + (4 r P_r + \tfrac{8}{3}) W
    = 0 , \nonumber\\ 
  \fl 2 \Kzr_{,r} + \tfrac{4}{3} U_{,z} + \tfrac{2}{3} r W_{,z}
    + (12 P_r + 4 R_r + 2 r^{-1}) \Kzr + (8 P_z + 2 R_z) U
    + 4 P_z r W = 0.\nonumber\\
\end{eqnarray}
The extrinsic curvature variables $\Kzr$, $U$ and $W$ are all
evolved using their time evolution equations \cite{Garfinkle2001}.

The slicing condition is solved in the form \eref{e:GDSlicingCondition},
and this is a Helmholtz equation with the ``right sign''
($c \leqslant 0$ in \eref{e:LinearHelmholtz};
note the square bracket in \eref{e:GDSlicingCondition} is non-negative).

In order to solve for the shift vector, additional derivatives are
taken of equations \eref{e:FirstOrderShift}, which combine to two
decoupled second-order equations,
\begin{eqnarray}
  \label{e:SecondOrderShift}
  \beta^r{}_{,rr} + \beta^r{}_{,zz} = 2 (\alpha \Kzr)_{,z} 
    - (\alpha U)_{,r} , \nonumber\\
  \beta^z{}_{,rr} + \beta^z{}_{,zz} = 2 (\alpha \Kzr)_{,r} 
    + (\alpha U)_{,z} .  
\end{eqnarray}
These are Poisson equations ($c = 0$ in \eref{e:LinearHelmholtz})
and do not cause any problems.

\subsection{A fully constrained scheme: 
  Choptuik \etal \cite{Choptuik2003}}

A similar formulation was developed by 
Choptuik \etal \cite{Choptuik2003}.
Their definition of the variables $\bar \sigma$ and $\psi$ 
differs slightly from our $S$ and $\psi$,
\begin{equation}
  \bar \sigma = -S, \qquad \psi_\mathrm{Ch}^4 = \psi^4 \rme^{2rS},
\end{equation}
where the subscript 'Ch' refers to \cite{Choptuik2003}.
This difference does not have any consequences on the properties of
the elliptic equations that we are concerned with here and so for the
sake of consistency we continue to use our convention (which agrees
with the one in \cite{Garfinkle2001}). As a result the equations
displayed below differ from those in \cite{Choptuik2003} in a minor way.

In the same way as Garfinkle and Duncan, Choptuik \etal also solve the
Hamiltonian constraint, which is again indefinite.

Unlike Garfinkle and Duncan, however, they also solve the momentum
constraints. This is done by replacing $\Kzr$ and $U$ with first
derivatives of the shift using the gauge conditions \eref{e:FirstOrderShift}.
The momentum constraints \eref{e:GDMomentumConstraints} now read
\begin{eqnarray}
  \fl \tfrac{2}{3} \beta^r{}_{,rr} + \beta^r{}_{,zz} + \third \beta^z{}_{,rz}
    + \tfrac{2}{3} \alpha r W_{,r} + \alpha (4 r P_r + \tfrac{8}{3}) W 
    \nonumber\\  
    + (6 P_z + 2 R_z - A_z) \beta_+ + (4 P_r + 2 R_r - \tfrac{2}{3}
    A_r) \beta_- = 0 ,\nonumber\\
  \fl -\third \beta^r{}_{,rz} + \beta^z{}_{,rr} + \tfrac{4}{3} \beta^z{}_{,rr}
    + \tfrac{2}{3} \alpha r W_{,z} + 4 \alpha P_z r W \nonumber\\
    + (6 P_r + 2 R_r + r^{-1} - A_r) \beta_+ - (8 P_z + 2 R_z 
    - \tfrac{4}{3} A_z) \beta_- = 0 .
\end{eqnarray}
The principal part of these two coupled equations is elliptic and so 
far there is no need for concern.
A problem arises however when equations \eref{e:FirstOrderShift} are
substituted in the slicing condition \eref{e:GDSlicingCondition},
\begin{eqnarray}
  \label{e:CSlicingCondition}
  \fl \alpha_{,rr} + \alpha_{,zz} + (2 P_r + r^{-1}) \alpha_{,r} + 2 P_z \alpha_{,z}
    -\psi^4 \rme^{2rS}  \alpha^{-1} [ \tfrac{2}{3} \beta_-^2 + \half
    \beta_+^2 ] \nonumber\\
    + \tfrac{2}{3} \psi^4 \rme^{2rS} \beta_- r W - \half \psi^4
    \rme^{2rS} \alpha (rW)^2 = 0.
\end{eqnarray}
The term containing the square bracket has the ``wrong sign'',
$p = -1$ and $c \leqslant 0$ in \eref{e:NonlinearHelmholtz},
\eref{e:NonlinearHelmholtzExample}.

\subsection{A new fully constrained scheme}
\label{s:ORscheme}

We observed that in both of the above schemes, the Hamiltonian
constraint was indefinite, and in the second one, the
slicing condition was, too. We now present a scheme in which both
equations and in fact all the elliptic equations that are being solved
are definite.

The Hamiltonian constraint can be cured by rescaling the extrinsic
curvature variables with a suitable power of the conformal factor,
\begin{equation}
  \label{e:ExtrinsicCurvatureRescaling}
  \{ \tKzr, \tU, \tW \} \equiv \psi^p \{ \Kzr, U, W \}.
\end{equation}
In terms of the new variables, the exponent of $\psi$ multiplying 
the second square bracket in \eref{e:GDHamiltonianConstraint} is
$5 - 2p$ so that the equation becomes definite for 
$p \geqslant 5/2$. There is a preferred choice: for $p = 6$ the terms
containing derivatives of $\psi$ in the momentum 
constraints \eref{e:GDMomentumConstraints} all
cancel under the substitution \eref{e:ExtrinsicCurvatureRescaling}.
The same rescaling of the extrinsic curvature was applied 
by Abrahams and Evans \cite{Abrahams1993}. 
Their scheme is however not fully constrained--the extrinsic curvature
variables are evolved as in \cite{Garfinkle2001}.

The indefiniteness of the slicing condition was caused by the
substitution \eref{e:FirstOrderShift}, more precisely by its $\alpha$
dependence. The original motivation for this substitution was the
desire to be able to solve the momentum constraints. However we can
still do this as before if we introduce a new vector $\eta^A$ and set
\begin{eqnarray}
   \label{e:EtaDefinition}
  \eta_+ \equiv \eta^z{}_{,r} + \eta^r{}_{,z} = 2 \tKzr, \nonumber\\
  \eta_- \equiv \eta^r{}_{,r} - \eta^z{}_{,z} = - \tU.
\end{eqnarray}
The momentum constraints are then solved for $\eta^A$. The price we
have to pay is that we still need to solve the spatial gauge conditions
\eref{e:SecondOrderShift}, where now $\Kzr$ and $U$ are expressed in
terms of $\eta_\pm$. That is, we have to solve two more elliptic
equations than Choptuik \etal.

Let us now write out all the elliptic equations explicitly.
The momentum constraints are
\begin{eqnarray}
  \label{e:ORMomentumConstraints}
  \fl \tfrac{2}{3} \eta^r{}_{,rr} + \eta^r{}_{,zz} + \third \eta^z{}_{,rz}
    + 2 R_r \eta_- + 2 R_z \eta_+ + \tfrac{2}{3} r \tW_{,r} 
    + \tfrac{8}{3} \tW = 0 , \nonumber\\
  \fl -\third \eta^r{}_{,rz} + \eta^z{}_{,rr} + \tfrac{4}{3} \eta^z{}_{,rr}
    - 2 R_z \eta_- + (2 R_r + r^{-1}) \eta_+ + \tfrac{2}{3} r \tW_{,z} = 0 .
\end{eqnarray}
The Hamiltonian constraint is
\begin{eqnarray}
  \label{e:ORHamiltonianConstraint}
  \fl \psi_{,rr} + \psi_{,zz} + r^{-1} \psi_{,r} + \quarter \psi 
    [ (rS)_{,rr} + (rS)_{,zz} ] \nonumber\\
    + \tfrac{1}{48} \psi^{-7} \rme^{2rS} [ (2 \eta_- - r \tW)^2
    + 3 (r \tW)^2 + 3 \eta_+^2 ] = 0 .
\end{eqnarray}
The slicing condition is
\begin{eqnarray}
  \label{e:ORSlicingCondition}
  \fl \alpha_{,rr} + \alpha_{,zz} + (2 P_r + r^{-1}) \alpha_{,r} + 2 P_z \alpha_{,z}
  \nonumber\\
  - \tfrac{1}{6} \alpha \psi^{-8} \rme^{2rS} [ (2 \eta_- - r \tW)^2
    + 3 (r \tW)^2 + 3 \eta_+^2 ] = 0 .
\end{eqnarray}
The spatial gauge conditions are
\begin{eqnarray}
  \label{e:ORSpatialGaugeConditions}
  \fl \beta^r{}_{,rr} + \beta^r{}_{,zz} - \alpha \psi^{-6} [
    \eta^r{}_{,rr} + \eta^r{}_{,zz} %\nonumber\\ 
    - (6 P_r - A_r) \eta_- - (6 P_z - A_z) \eta_+ ] = 0 , \nonumber\\
  \fl \beta^z{}_{,rr} + \beta^z{}_{,zz} - \alpha \psi^{-6} [
    \eta^z{}_{,rr} + \eta^z{}_{,zz} %\nonumber\\ 
    + (6 P_z - A_z) \eta_- - (6 P_r - A_r) \eta_+ ] = 0 .
\end{eqnarray}
We note that \eref{e:ORMomentumConstraints}--\eref{e:ORSpatialGaugeConditions}
form a hierarchy: the equations are successively solved for $\eta^A$, $\psi$,
$\alpha$, and $\beta^A$. After substituting the solutions of the
previous equations, each equation in the hierarchy can be regarded as a
decoupled scalar elliptic equation, or elliptic system in the case of  
\eref{e:ORMomentumConstraints}.
The terms in the second lines of \eref{e:ORHamiltonianConstraint} and
\eref{e:ORSlicingCondition} now have the ``right signs''.
An exception common to all the schemes discussed in this section 
is the term multiplying $\psi$ in the first line of 
\eref{e:ORHamiltonianConstraint}--in general one expects $S$ to oscillate 
so that $\Delta (rS)$ can have either sign. This is the usual
difficulty one faces in conformal formulations of the initial value
equations, see section \ref{s:CTS}.

The variable $S$ and its ``time derivative'' $\tW$ are evolved. This
pair of evolution equations corresponds to the one dynamical degree of
freedom. Note that if we had not restricted the Killing vector to be
hypersurface orthogonal then there would be a second dynamical degree
of freedom. In linearized theory these two degrees of freedom can be
understood as the two polarization states of a gravitational wave.
There are additional evolution equations for $\psi$, $\eta_+ = 2 \tKzr$ 
and $\eta_- = -\tU$ that are not actively enforced but that can be used
in order to test the accuracy of a numerical implementation.
All the evolution equations are given in \ref{s:EvolutionEquations}.
Here we remark that assuming a solution to the elliptic equations is
given, the principal part of the evolution equations is that of a wave
equation,
\begin{equation}
  \label{e:WaveEvolutionEquation}
  \left[ \alpha^{-1}(\partial_t - \beta^A \partial_A) \right]^2 S
  \simeq \psi^{-4} \rme^{-2rS} (S_{,rr} + S_{,zz}),
\end{equation}
where $\simeq$ denotes equality to principal parts.
This equation is clearly hyperbolic, a necessary criterion for the
well posedness of the Cauchy problem.
See also \cite{CorderoCarrion2008} for a recent analysis of the hyperbolic
part of a fully $3+1$ dimensional constrained evolution scheme based
on the Dirac gauge.

\subsection{Relation to the (extended) conformal thin sandwich formulation}
\label{s:CTS}

Our discussion of different constrained evolution schemes for the
axisymmetric Einstein equations is closely related to conformal
approaches to solving the initial value equations in standard $3+1$
dimensional spacetime. Here one seeks to find a spatial metric 
$\gamma_{ij}$ and extrinsic curvature $K_{ij}$ satisfying the 
constraint equations on the initial slice, and often also a lapse 
$\alpha$ and shift $\beta^i$ satisfying suitable gauge conditions.
This is done by setting
\begin{equation}
  \gamma_{ij} = \psi^4 \tilde \gamma_{ij} ,
\end{equation}
where $\psi$ is the conformal factor, and the conformal metric 
$\tilde \gamma_{ij}$ is assumed to be given.
For simplicity and for analogy with the axisymmetric formulations
discussed above, we impose the gauge condition 
$\partial_t \tilde \gamma_{ij} = 0$. (In the axisymmetric case, we
only controlled the $r, z$ components, $\partial_t \tilde \gamma_{AB} = 0$.)
We also assume maximal slicing $K = 0$ throughout, and we work in vacuum.

It is well known in the conformal approach that the extrinsic
curvature $K_{ij}$ cannot be freely specified \cite{York1979}; 
instead it has to be conformally rescaled,
\begin{equation}
  \label{e:CTTExtrinsicCurvatureDecomposition}
  K_{ij} - \third K \gamma_{ij} = \psi^{-10} \tilde A_{ij}.
\end{equation}
This corresponds to the proposed rescaling of the extrinsic curvature 
variables \eref{e:ExtrinsicCurvatureRescaling} in the new 
axisymmetric scheme.
The Hamiltonian constraint now takes the form of the
Lichnerowicz equation,
\begin{equation}
  \label{e:CTTHamiltonianConstraint}
  8 \tilde \Delta \psi - \tilde R \psi + \psi^{-7} \tilde A_{ij}
  \tilde A^{ij} = 0,
\end{equation}
where $\tilde \Delta$ is the covariant Laplacian and $\tilde R$ the Ricci
scalar of the conformal metric $\tilde \gamma_{ij}$.
Note again that the last term in \eref{e:CTTHamiltonianConstraint} 
has the ``right sign'' for linearization stability, 
cf.~\eref{e:ORHamiltonianConstraint}. As pointed out in the previous
subsection, the linear term $-\tilde R \psi$ can have either sign.
However, $\tilde R < 0$ does not \emph{necessarily} imply that
multiple solutions exist.  There is a well-developed theory for
existence and uniqueness of solutions to \eref{e:CTTHamiltonianConstraint},
see \cite{Brill1981}.

In order to solve the momentum constraints, York's original \emph{conformal
transverse traceless} (CTT) method introduces a vector $\eta^i$ and sets
\begin{equation}
  \label{e:CTTExtrinsicCurvatureDecomposition2}
  \tilde A_{ij} =(\mathbb{L} \eta)_{ij},
\end{equation}
where $\mathbb{L}$ is the conformal Killing operator defined as
\begin{equation}
  (\mathbb{L} \eta)_{ij} = 2 \tilde \nabla_{(i} \eta_{j)} -
  \tfrac{2}{3} \tilde \gamma_{ij} \tilde \nabla_k \eta^k .
\end{equation}
The momentum constraints now read
\begin{equation}
  \tilde \nabla_j (\mathbb{L} \eta)^{ij} = 0,
\end{equation}
in analogy with \eref{e:ORMomentumConstraints}.

In the CTT approach, any gauge conditions are solved \emph{after} a
solution of the constraint equations has been found. 
For example, maximal slicing $K = \partial_t K = 0$ implies
the following elliptic equation for the conformal lapse $\tilde \alpha =
\psi^{-6} \alpha$, 
\begin{equation}
  \label{e:CTTSlicingCondition}
  \tilde \Delta \tilde \alpha - \tilde A_{ij} \tilde A^{ij} 
    \tilde \alpha = 0,
\end{equation}
where \eref{e:CTTExtrinsicCurvatureDecomposition2} is substituted.
This equation has the ``right sign'', as it has in the new
axisymmetric formulation \eref{e:ORSlicingCondition} and in the one by
Garfinkle and Duncan \eref{e:GDSlicingCondition}.

In contrast, the \emph{extended conformal thin sandwich} (XCTS)
method \cite{Pfeiffer2003}
directly expresses the extrinsic curvature in terms of the shift $\beta^i$,
\begin{equation}
  \tilde A_{ij} = \frac{(\mathbb{L} \beta)_{ij}}{2 \tilde \alpha}
\end{equation}
instead of \eref{e:CTTExtrinsicCurvatureDecomposition2}.
As a result, the slicing condition \eref{e:CTTSlicingCondition}
acquires the wrong sign. 
This is precisely what happens in the scheme by Choptuik \etal, 
cf.~\eref{e:FirstOrderShift} and \eref{e:CSlicingCondition}.

Remarkably, a numerical study of the XCTS equations \cite{Pfeiffer2005}
showed that this system does admit nonunique solutions. Two solutions
were found for small perturbations of Minkowski space, one of them
even containing a black hole. The two branches meet for a certain
critical amplitude of the perturbation. This parabolic branching
was explained using Lyapunov-Schmidt theory in \cite{Walsh2007}.
Because of the similarity with the XCTS equations, it is conceivable
that the constrained axisymmetric formulation of Choptuik \etal 
\cite{Choptuik2003} might show a similar branching behaviour. 
This is clearly undesirable for numerical evolutions because the
elliptic solver might jump from one solution branch to the other
during the course of an evolution. However even before this can happen the
multigrid method used in \cite{Choptuik2003} will fail to converge, as
explained in section \ref{s:Ell}.

%%%%%%%%%%%%%%%%%%%%%%%%%%%%%%%%%%%%%%%%%%%%%%%%%%%%%%%%%%%%%%%%%%%%%%%%%%%%%%%

\section{Numerical method}
\label{s:NumMethod}

In this section we describe a numerical implementation of the new
fully constrained scheme presented in section \ref{s:ORscheme}.

The equations are discretized using second-order accurate finite
differences in space. A collection of the finite difference operators
we use can be found in \ref{s:Discretization}.
Similarly to \cite{Garfinkle2001}, and unlike \cite{Choptuik2003},
we use a \emph{cell-centred} grid to cover the spatial domain
$[0, r_\mathrm{max}] \times [0, z_\mathrm{max}]$:
grid points are placed at coordinates $r_i = (i - \half) \Delta r$, 
$1 \leqslant i \leqslant N_r$,
where $\Delta r = r_\mathrm{max} / N_r$ is the grid spacing and 
$N_r$ is the number of grid points in the $r$ direction. 
(Corresponding relations hold in the $z$ direction.)
Note that no grid points are placed at the boundaries.
\emph{Ghost points} are placed at $r_0 = -\half \Delta r$ and 
at $r_{N_r + 1} = r_\mathrm{max} + \half \Delta r$.
The values at these ghost points are set according to the boundary
conditions, as described in the following. 
Here we only refer to the ``physical'' grid boundaries at $r=0$, $z=0$,
$r=r_\mathrm{max}$ and $z=z_\mathrm{max}$. In the adaptive mesh
refinement approach discussed further below, additional finer
grids are added that do not cover the entire spatial domain.
These finer grids are also surrounded by ghost points. On grid boundaries 
that do not coincide with a ``physical'' boundary, ghost point values
are interpolated from the coarser grid \cite{Berger1984}.

The boundary conditions at $r = 0$ follow from regularity on axis 
(see \cite{Rinne2005} for a rigorous discussion): 
either a Dirichlet or a Neumann condition
is imposed depending on whether the variable is an odd or even
function of $r$. All the equations being solved (both the elliptic
equations and the evolution equations) are regular on the axis provided
that these boundary conditions are satisfied.
In addition, we impose reflection symmetry about $z = 0$ so that the
variables are either odd or even functions of $z$, and this implies
Dirichlet or Neumann conditions at $z = 0$.
The $r$ and $z$ parities of all the variables are listed in 
\ref{s:Discretization}.

At the outer boundaries $r = r_\mathrm{max}$ and $z = z_\mathrm{max}$,
we impose Dirichlet conditions on the gauge variables,
$\alpha = 1$ and $\beta^A = 0$.
For the variables $\{ \psi, \eta^A \} \ni u$, we impose
\begin{equation}
  \label{e:FalloffBC}
  u = u_\infty + \tfrac{f(\theta)}{R},
\end{equation}
where $R = \sqrt{r^2 + z^2}$ and $\theta = \tan(r/z)$ are spherical
polar coordinates and $u_\infty$ is the value of $u$ at spacelike infinity,
i.e., $\psi_\infty = 1$ and $\eta^A_\infty = 0$.
This boundary condition obviously holds up to terms of $O(R)^{-2}$ for any
asymptotically flat solution of the constraint equations.
For the ``dynamical'' fields $\{ S, \tW \} \ni u$, we follow 
\cite{Choptuik2003} and impose a Sommerfeld condition at the outer boundary,
\begin{equation}
  \label{e:SommerfeldBC}
  (\partial_t + \partial_R) [R (u - u_\infty)] = 0 .
\end{equation}
This condition is only exact for the scalar wave equation in spherical
symmetry. It is however expected to be a reasonable first
approximation because $S$ and $\tW$ obey a wave equation 
\eref{e:WaveEvolutionEquation} to principal parts and the elliptic
variables will be close to their flat-space values near the outer
boundary ($\alpha = \psi = 1$, $\beta^A = \eta^A = 0$).
See \ref{s:Discretization} for details on the discretization at the
outer boundary.

The evolution equations are integrated forward in time using the
Crank-Nicholson method; this method is second-order accurate in time.
The resulting implicit equations are solved
by an outer Gauss-Seidel relaxation (in red-black ordering),
and an inner Newton-Raphson method in order to solve for the vector 
of unknowns at each grid point.
A typical value of the CFL number $\lambda \equiv \Delta t / \min(\Delta
r, \Delta z)$ we use is $\lambda = 0.5$.
Fourth-order Kreiss-Oliger dissipation \cite{Kreiss1973} is added 
to the right-hand sides of the evolution equations, with a typical 
parameter value of $\epsilon = 0.5$.

The elliptic equations are solved using a multigrid method \cite{Brandt1977}.
The Full Approximation Storage (FAS) variant of the method
enables us to solve nonlinear equations directly, i.e.~we do
\emph{not} apply an outer Newton-Raphson iteration in order to
obtain a sequence of linear problems. In the relaxation step
of the multigrid algorithm, a nonlinear Gauss-Seidel relaxation 
(in red-black ordering) is directly applied to the full nonlinear
equations. At each grid point, we solve simultaneously for the
unknowns $\eta^A$, $\psi$, $\alpha$, and $\beta^A$. Only the
Hamiltonian constraint \eref{e:ORHamiltonianConstraint} requires 
the solution of a (scalar) nonlinear equation, and this is done using
Newton's method; a single iteration is found to be sufficient. 
All interior grid points are relaxed and afterwards the values at 
the ghost points are filled according to the boundary conditions.
In order to transfer the numerical solution between the grids, 
we use biquadratic interpolation for prolongation and linear averaging 
for restriction. 

For the prolate wave evolved in section \ref{s:ProlateWave}, the elliptic
equations become highly anisotropic and the standard pointwise
Gauss-Seidel relaxation employed in the multigrid method no
longer converges.  A common cure to this problem is line 
relaxation \cite{Trottenberg2001}. 
We solve for the unknowns at all grid points in a line $z = \mathrm{const}$
simultaneously. One Newton-Raphson step is applied to treat the
nonlinearity and the resulting tridiagonal linear system is solved
using the Thomas algorithm. Note that this method has the same computational 
complexity as the pointwise Gauss-Seidel relaxation. 

The wide range of length scales in the solutions we are interested in
necessitates a position-dependent grid resolution. The classic 
adaptive mesh refinement (AMR) algorithm by Berger and 
Oliger \cite{Berger1984} was designed for hyperbolic
equations. Including elliptic equations in this approach is rather
complicated. A solution with numerical relativity applications in mind
was suggested by Pretorius and Choptuik \cite{Pretorius2006a},
and we shall use their algorithm here, with minor modifications due to
the fact that our grids are cell centred rather than vertex centred.
The key idea of the algorithm is that solution of the elliptic
equations on coarse grids is deferred until all finer grids have
reached the same time; meanwhile the elliptic unknowns are linearly
extrapolated in time and only the evolution equations are solved.
We have found that this approach works well as long as no grid boundaries
are placed in the highly nonlinear region. In particular, adaptive
generation of finer grids in the course of the evolution causes
small but noticeable reflections that from our experience make the
study of problems such as Brill wave critical collapse unfeasible.
For this reason, the evolutions presented in this article use
\emph{fixed} mesh refinement (FMR), i.e.~the grid hierarchy is defined at
the beginning of the simulation and remains unchanged as time evolves.

Finally we briefly discuss how an apparent horizon is found in a 
$t = \mathrm{const}$ slice. 
The horizon is parametrized as a curve $R = f(\theta)$
in spherical polar coordinates. Requiring the expansion of the
outgoing null rays emanating from the horizon to vanish yields a
second order ordinary differential equation, which is solved using the
shooting method. The boundary conditions are $f'(0) = f'(\pi/2) = 0$,
i.e., the horizon has no cusps on the axes. We follow an idea of
Garfinkle and Duncan \cite{Garfinkle2001} in order to monitor the 
approach to apparent horizon formation. For each point on the axis 
$r = 0 \Leftrightarrow \theta = 0$, we find the angle $\gamma$ 
at which the curve starting from that point meets the 
$z = 0 \Leftrightarrow \theta = \pi/2$ axis, 
\begin{equation}
  \label{e:GammaAngle}
  \cot \gamma = \left. \frac{f'}{f} \right\vert_{\theta = \pi/2}.
\end{equation}
We find the maximum $\gamma_\mathrm{max}$ of this angle over all such
curves. Obviously $\gamma_\mathrm{max} = \pi/2$ for an apparent horizon, 
and the deviation from this value indicates how close we are to the
formation of an apparent horizon.

%%%%%%%%%%%%%%%%%%%%%%%%%%%%%%%%%%%%%%%%%%%%%%%%%%%%%%%%%%%%%%%%%%%%%%%%%%%%%%%

\section{Numerical results}
\label{s:NumResults}

As an application of our numerical implementation, we consider vacuum
axisymmetric gravitational waves, so-called Brill waves \cite{Brill1981}.
The initial slice is taken to be time-symmetric so that
$\tilde W = \eta^A = 0$ initially.
We consider the same initial data for the function $S$ as 
in \cite{Abrahams1992} and \cite{Garfinkle2001},
\begin{equation}
  \label{e:InitialData}
  S = a \, r \exp \left( 
      -\frac{r^2}{\sigma_r^2} - \frac{z^2}{\sigma_z^2} \right),
\end{equation}
where $a$, $\sigma_r$ and $\sigma_z$ are constants. The initial lapse and
shift are taken to be $\alpha = 1$ and $\beta^A = 0$.
The momentum constraints (\ref{e:ORMomentumConstraints}) are trivially 
satisfied initially and only the Hamiltonian constraint 
(\ref{e:ORHamiltonianConstraint}) needs to be solved.

\subsection{Convergence test}
\label{s:ConvTest}

\begin{figure}
  \bigskip\medskip
  \begin{minipage}[t]{0.5\textwidth}
    \includegraphics[width=0.85\textwidth]{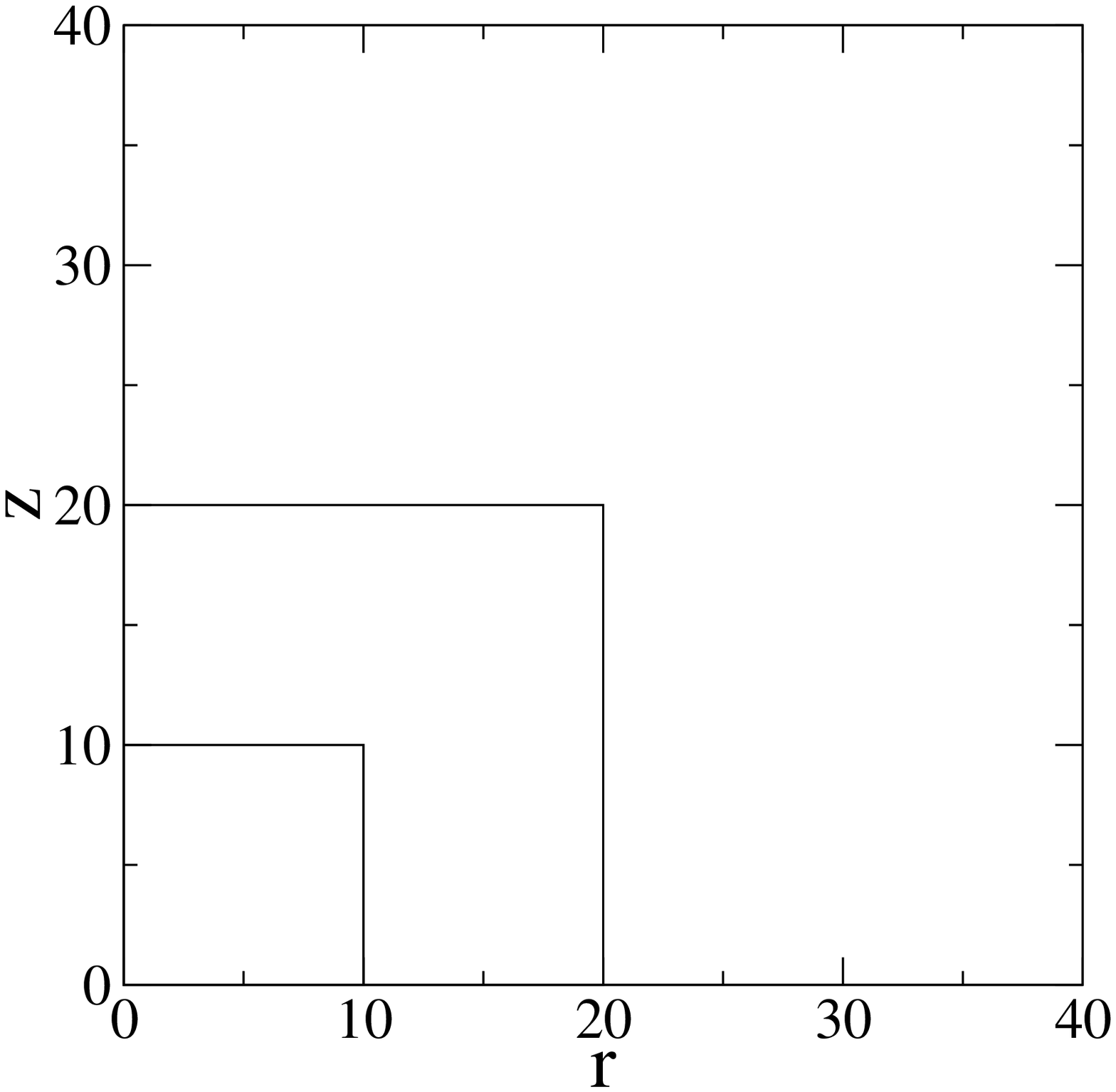}
  \end{minipage}
  \begin{minipage}[t]{0.5\textwidth}
    \includegraphics[width=0.85\textwidth]{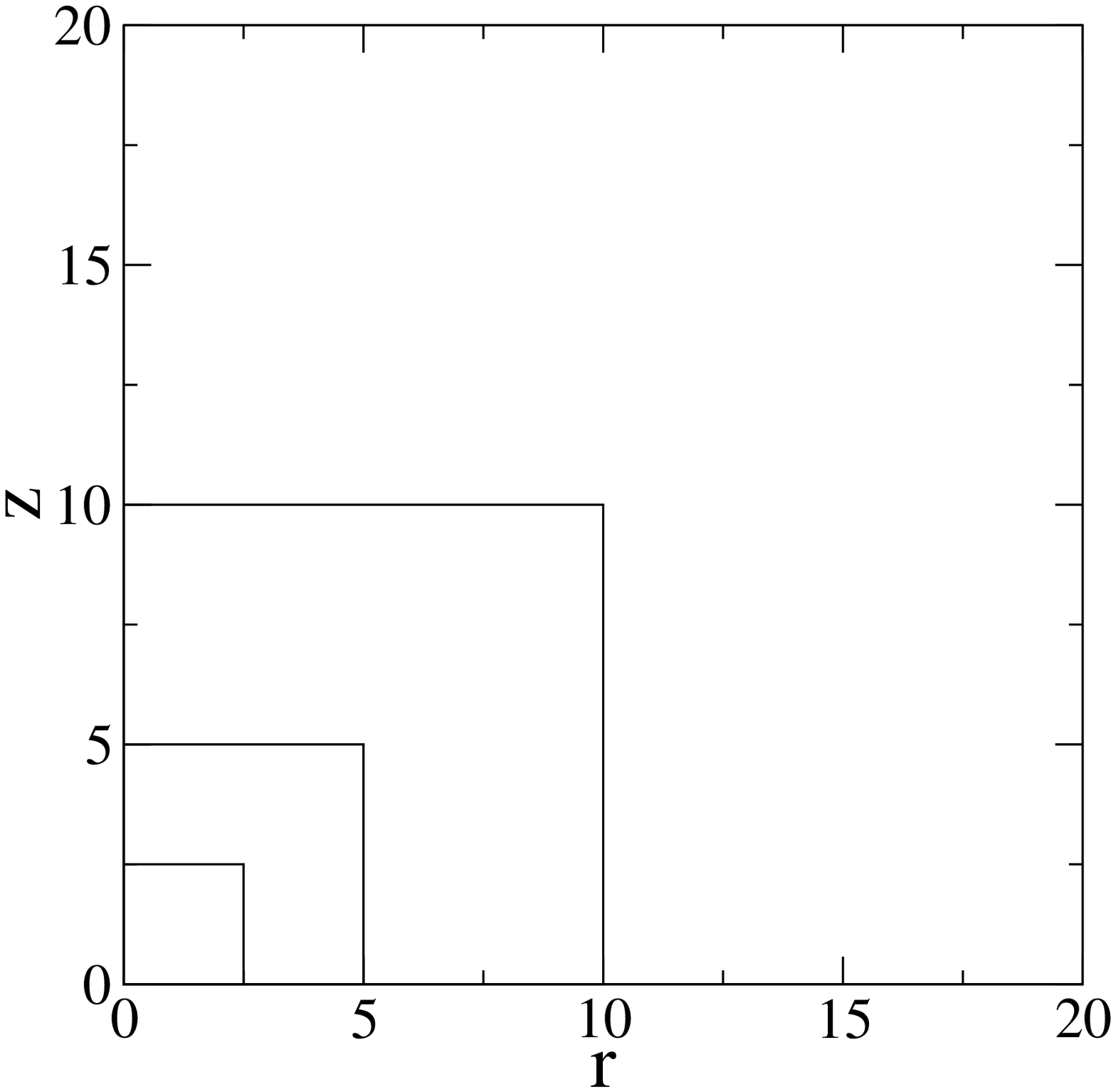}
    \vspace{0.8cm}\\
    \includegraphics[width=0.85\textwidth]{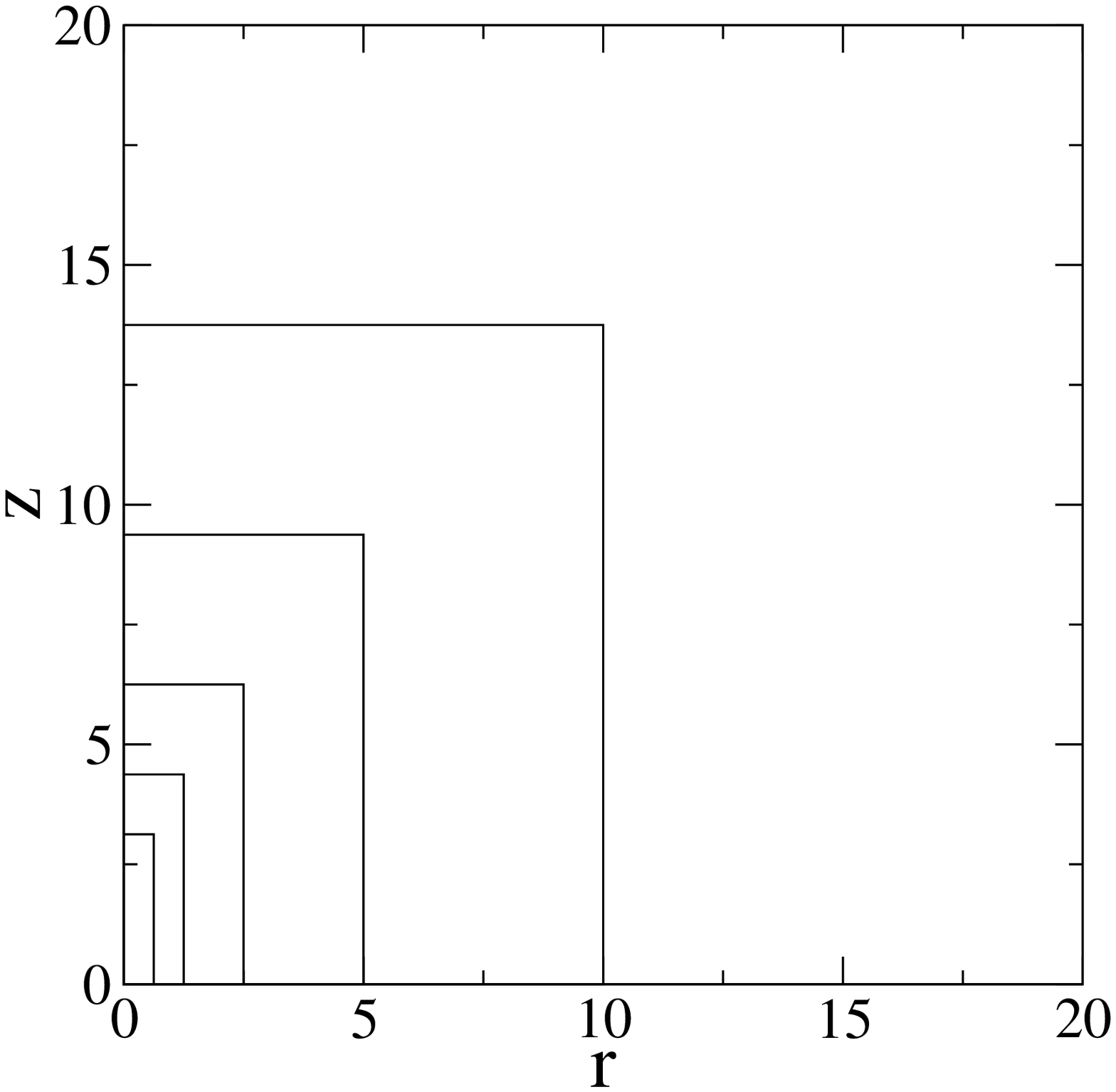}
  \end{minipage}
  \caption{\label{f:FMRhierarchy}
    FMR grid hierarchies used for the Brill wave evolutions presented
    in this paper.
    Top left: $\sigma_r = \sigma_z = a = 1$ (section \ref{s:ConvTest}); 
    top right: $\sigma_r = \sigma_z = 1$, $a = 8.5$ 
    (section \ref{s:SpherWave}),
    bottom: $\sigma_r = 0.128$, $\sigma_z = 1.6$, $a = 325$
    (section \ref{s:ProlateWave}).
  }
\end{figure}

In order to check convergence of the code, we first consider a
wave with parameters $a = \sigma_r = \sigma_z = 1$. 
This will disperse rather than collapse to a black hole but is still
well in the nonlinear regime.
The ADM mass is $M_\mathrm{ADM} =0.034$. 
We take the domain size to be $r_\mathrm{max} = z_\mathrm{max} = 40$.
The FMR hierarchy contains three grids (figure \ref{f:FMRhierarchy}).
All the grids contain the origin, are successively refined by a factor
of $2$, and all have the same number of grid points $N_r=N_z$.

\begin{figure}
  \begin{minipage}[t]{0.5\textwidth}
    \includegraphics[width=0.95\textwidth]{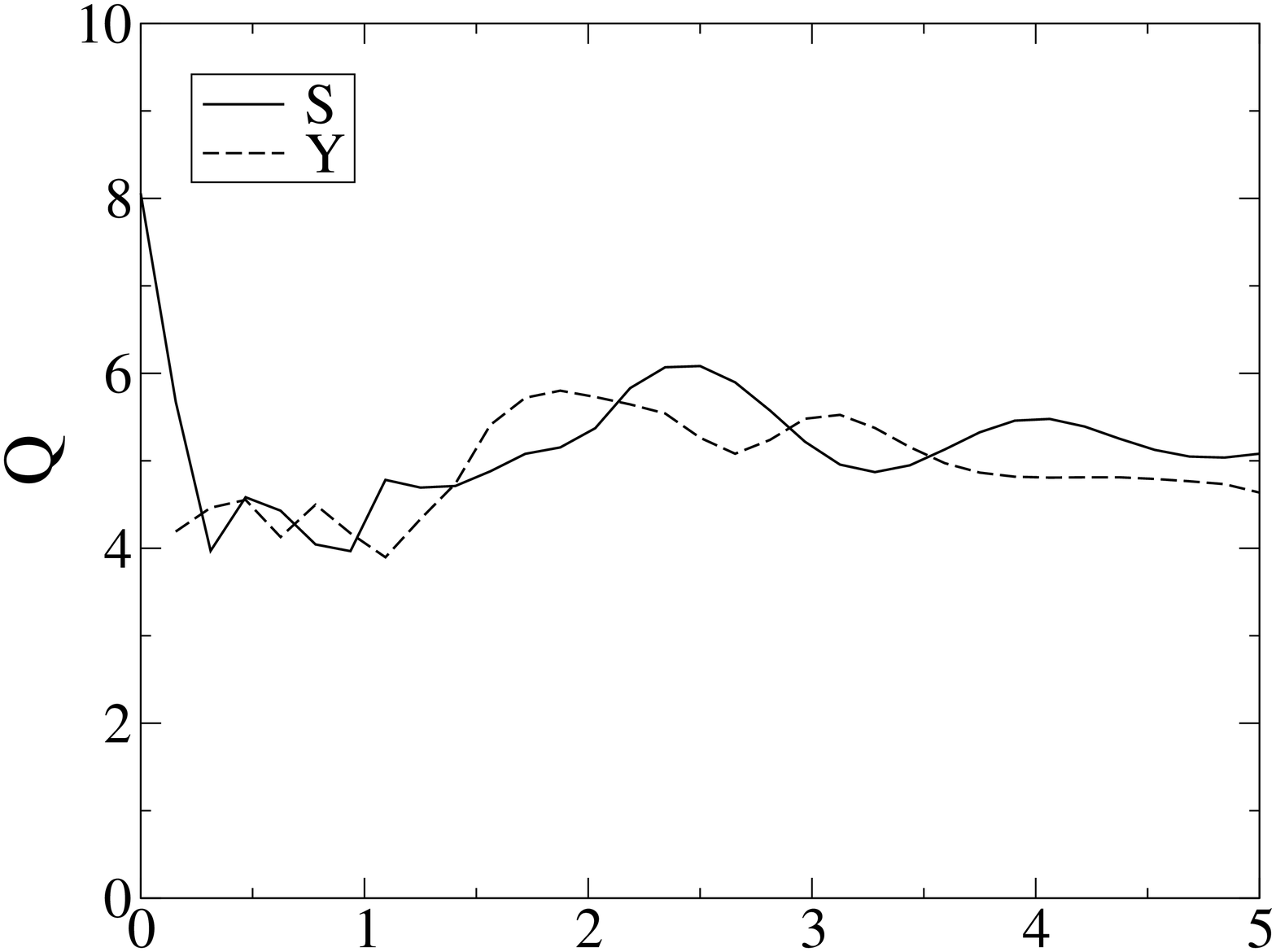}
    \medskip\\
    \includegraphics[width=0.95\textwidth]{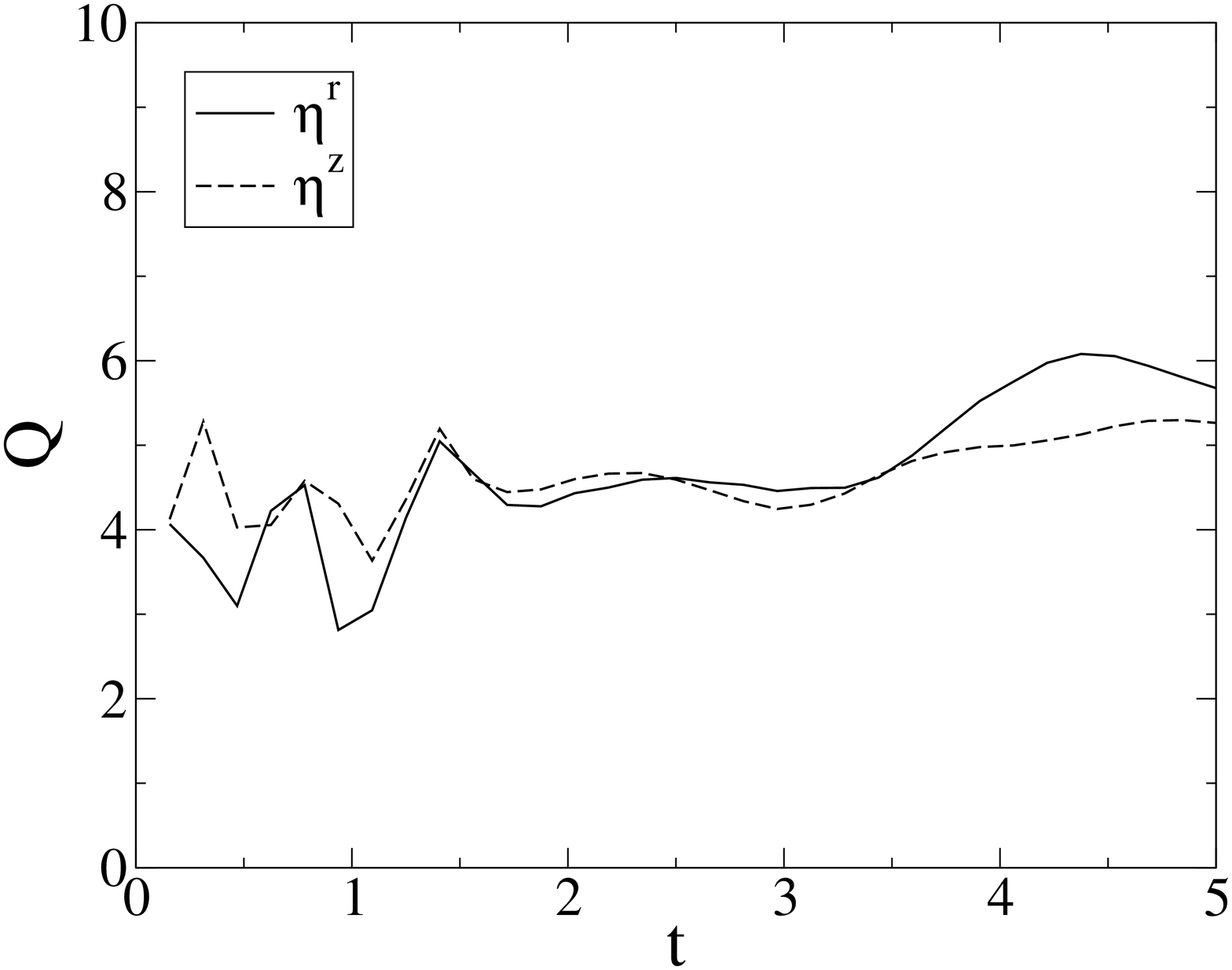}
  \end{minipage}
  \begin{minipage}[t]{0.5\textwidth}
    \includegraphics[width=0.95\textwidth]{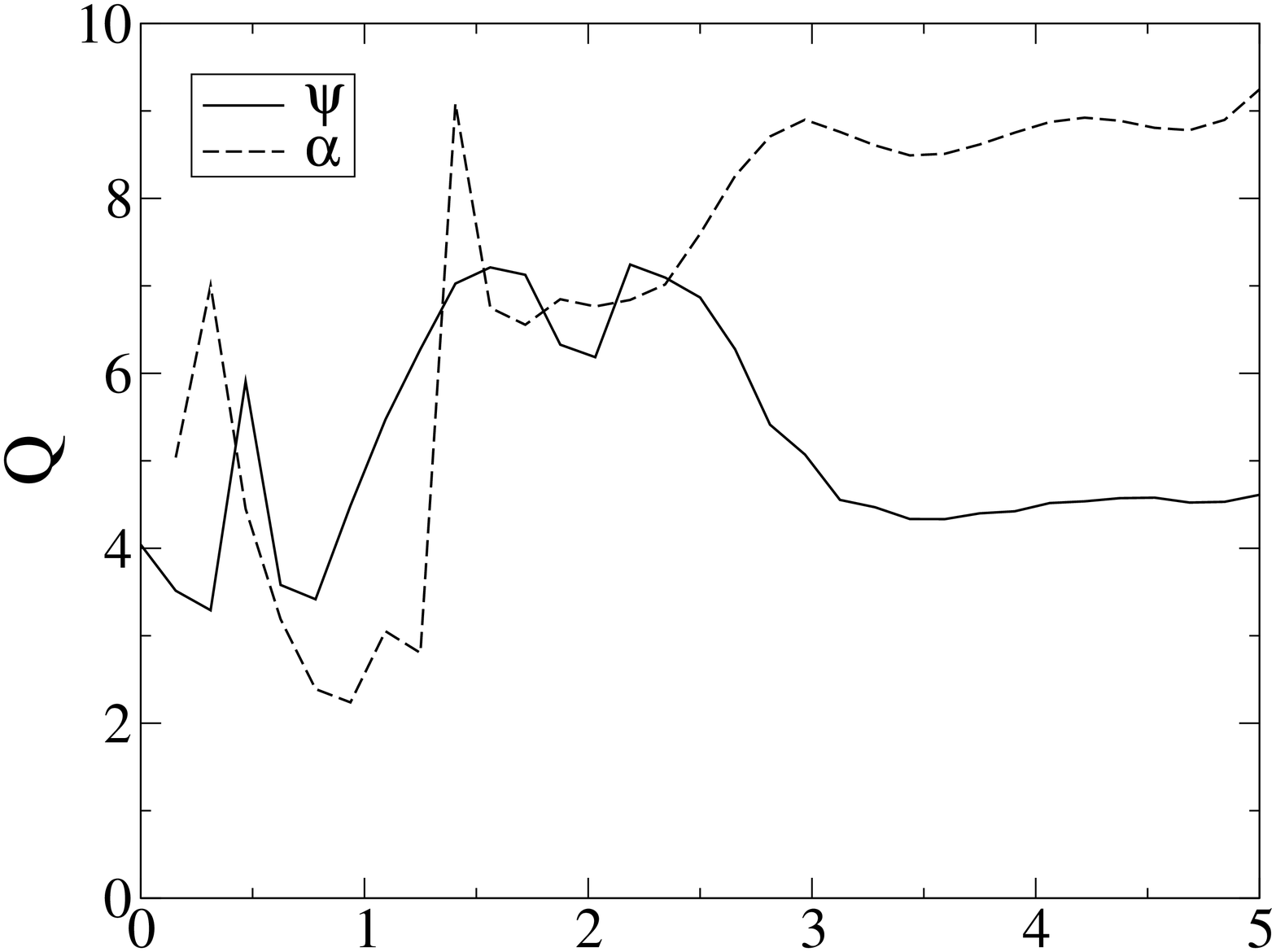}
    \medskip\\
    \includegraphics[width=0.95\textwidth]{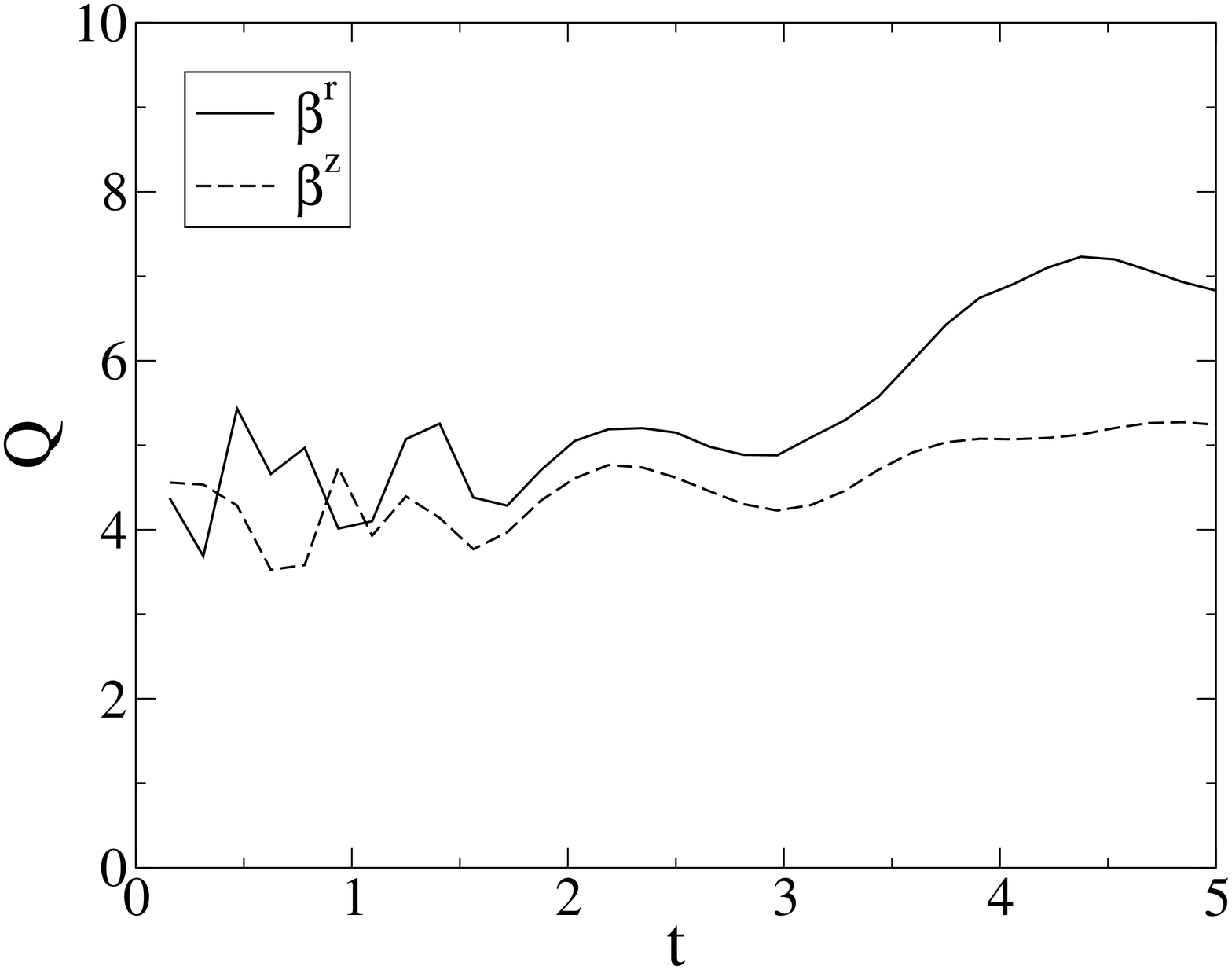}
  \end{minipage}
  \caption{\label{f:1.0conv}
    Three-grid convergence factors \eref{e:Q} for a Brill wave with 
    $a = \sigma_r = \sigma_z = 1$ computed from the three
    resolutions $N_r = N_z = 64, \, 128, \, 256$.
  }
\end{figure}

We run the simulation with three different resolutions, 
$N_r=N_z \in \{64,128,256\}$.
This enables us to carry out a three-grid convergence test: for each
variable $u$ we define a convergence factor
\begin{equation}
  \label{e:Q}
  Q_u \equiv \frac{\parallel u_{4h} - u_{2h} \parallel}
    {\parallel u_{2h} - u_h \parallel},
\end{equation}
with the indices referring to the grid spacing.
The norms are discrete $L^2$ norms taken over the subsets of all grids
in the FMR hierarchy that do not overlap with finer grids.
For a second-order accurate numerical method we expect $Q_u = 4$.
Figure \ref{f:1.0conv} confirms that the code is approximately
second-order convergent. (Occasional values $Q_u > 4$ are not uncommon 
in similar numerical schemes \cite{Choptuik2003,Pretorius2006a}.)

\begin{figure}
  \begin{minipage}[t]{0.5\textwidth}
    \includegraphics[width=0.95\textwidth]{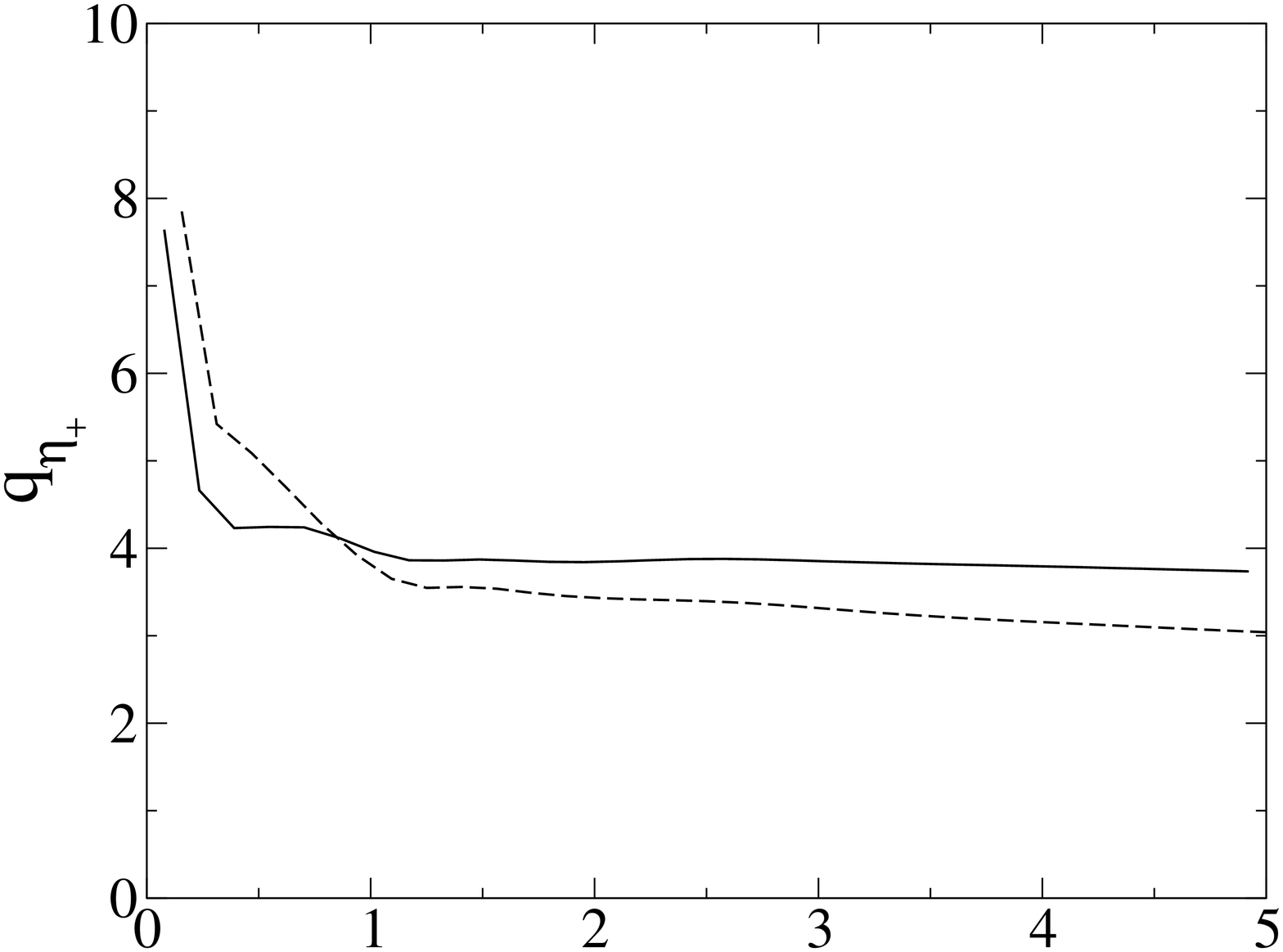}
    \medskip\\
    \includegraphics[width=0.95\textwidth]{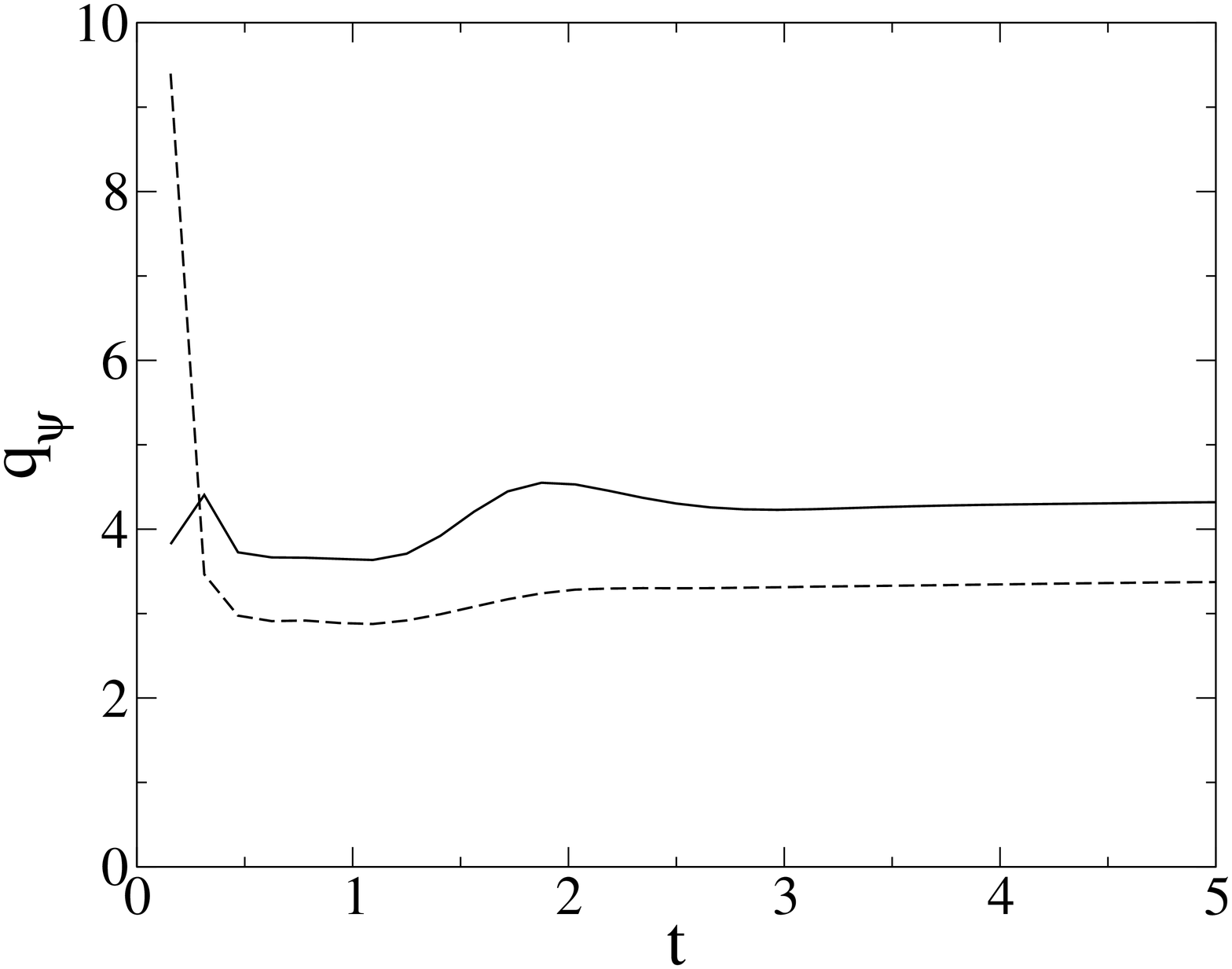}
  \end{minipage}
  \begin{minipage}[t]{0.5\textwidth}
    \includegraphics[width=0.95\textwidth]{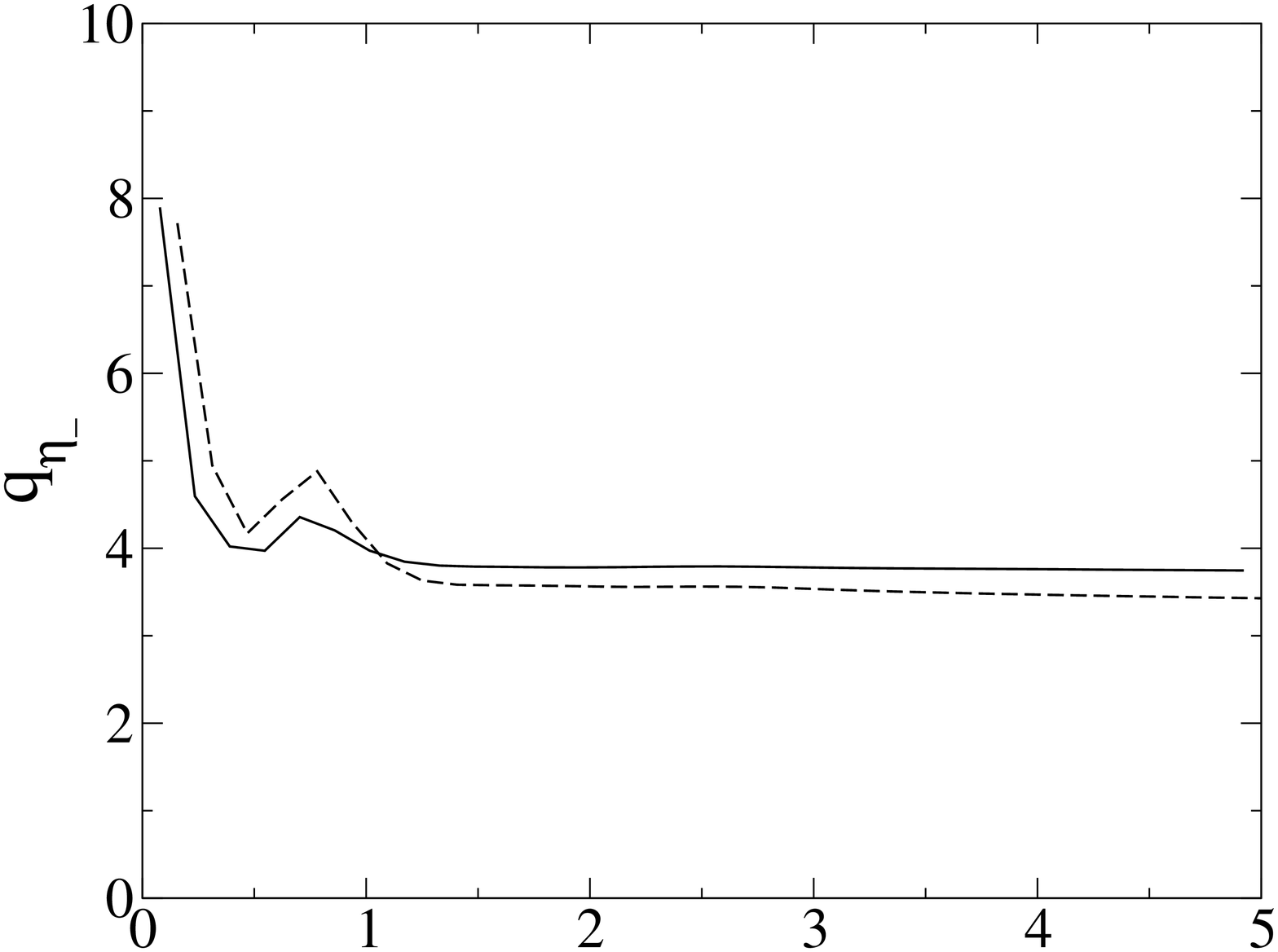}
    \medskip\\
    \includegraphics[width=0.95\textwidth]{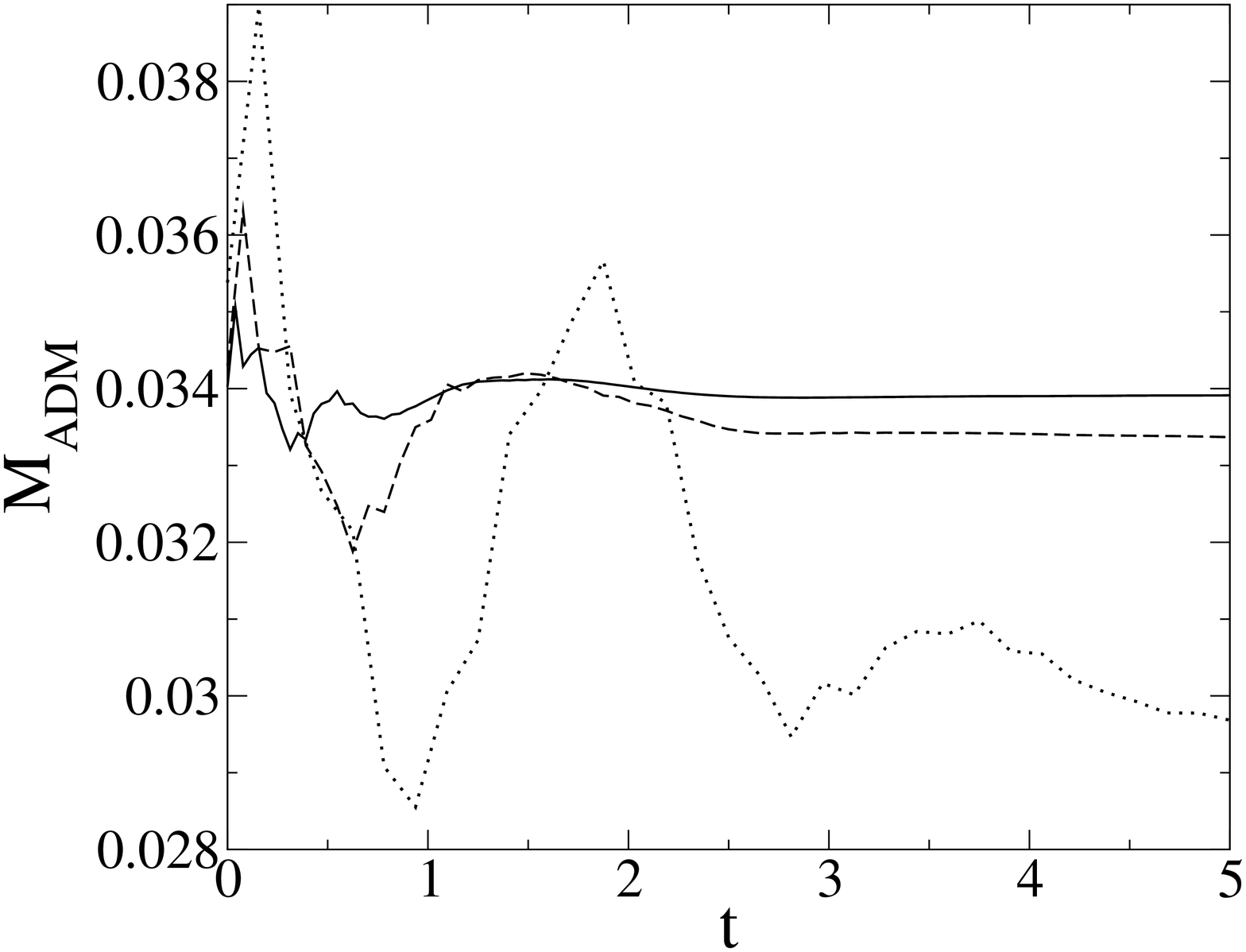}
  \end{minipage}
  \caption{\label{f:1.0resconv}
    Residual convergence factors \eref{e:q} for a Brill wave with 
    $a = \sigma_r = \sigma_z = 1$ computed from two pairs
    of resolutions, $N_r = N_z = 64:128$ (dashed) and $128:256$ (solid).
    The bottom right panel shows the numerically computed ADM mass 
    for the three different resolutions, $N_r = N_z = 64$ (dotted), 
    $128$ (dashed) and $256$ (solid).
  }
\end{figure}

As noted earlier, there are additional evolution equations for the
variables $\eta_+$, $\eta_-$ and $\psi$ that are not actively evolved
in our constrained evolution scheme. 
We use these to check the accuracy of the numerical implementation in
the following way.
We keep a set of auxiliary variables $\hat \eta_+$, $\hat \eta_-$ 
and $\hat \psi$ which are copied from their unhatted counterparts 
initially but evolved using the evolution equations 
\eref{e:psievolution}--\eref{e:etamevolution}.
During the evolution, we form the differences between the two sets.
Doing so for two different resolutions (grid spacings $h$ and $2h$)
allows us to define another convergence factor for each 
$u \in \{\eta_+, \eta_-, \psi\}$ (referred to as \emph{residual
  convergence} in the following),
\begin{equation}
  \label{e:q}
  q_u = \frac{\parallel\hat u_{2h} - u_{2h} \parallel}
  {\parallel \hat u_h - u_h \parallel} .
\end{equation}
The results in figure \ref{f:1.0resconv} are again compatible with
second-order convergence.

We note that the residual convergence test just presented is more
severe than the three-grid convergence test in the following sense.
For the residuals of the unsolved evolution equations to converge as 
desired, not only must the numerical solution be second-order convergent 
but the constraint and evolution equations 
\emph{and their boundary conditions} must be compatible.
No exact boundary conditions are known at a finite distance from the
source and compatibility of the boundary conditions we use is only 
achieved at infinity. 
We deliberately chose the domain size in this convergence test to be 
sufficiently large ($\sim 10^3 M_\mathrm{ADM}$) so that the effect of
the boundary on the convergence factors $q_u$ is small. 

As another consistency test, we compute a numerical approximation to
the ADM mass. 
This is evaluated as a surface integral on a sphere close 
to the outer boundary, at spherical radius $R = R_M$, 
\begin{equation}
  \label{e:ADMmass}
  M_\mathrm{ADM} = -R_M^2 \int_0^{\pi/2} J_A n^A \, 
    \sin \theta \, \rmd \theta, 
\end{equation}
where
\begin{equation}
  J_A = 2 \psi_{,A} + \half r S_{,A},
\end{equation}
$n_A$ is the unit normal in the spherical radial direction, and 
these expressions are valid in linearized theory.
We evaluate $M_\mathrm{ADM}$ in \eref {e:ADMmass} for two radii 
$R_M \in \{14,18\}$ and extrapolate to infinity assuming that
$M_\mathrm{ADM}(R_M) = M_\mathrm{ADM}(\infty) + \mathrm{const}/R_M$.
The result in figure \ref{f:1.0resconv} shows how numerical
conservation of the ADM mass improves with increasing resolution.

\subsection{``Spherical'' collapse}
\label{s:SpherWave}

Next we consider a wave with $\sigma_r = \sigma_z = 1$ and $a = 8.5$.
We refer to this as ``spherical'' because $\sigma_r = \sigma_z$,
although of course the actual wave is not spherically symmetric.
Unlike the one in section \ref{s:ConvTest}, this wave is
super-critical and will collapse to form a black hole.
The ADM mass is $M_\mathrm{ADM} \approx 2$.
We run the simulation for two different domain sizes, $r_\mathrm{max}
= z_\mathrm{max} \in \{ 10, 20 \}.$ 
The FMR hierarchy is of a similar type as in section \ref{s:ConvTest}.
On the smaller domain there are three grids and for the larger domain 
we add on another coarse grid (figure \ref{f:FMRhierarchy}).
We run the simulation with two different resolutions, $N_r = N_z \in
\{ 128, 256 \}$. 
In \cite{Garfinkle2001}, the same initial data were evolved on a
non-uniform grid with spacing $\Delta r = \Delta z = 1.92\times 10^{-2}$ 
close to the origin. 
This is comparable to our lower resolution grid hierarchy, which has 
grid spacing $\Delta r = \Delta z = 1.95\times 10^{-2}$ on the finest grid.

\begin{figure}
  \begin{minipage}[t]{0.5\textwidth}
    \includegraphics[width=0.95\textwidth]{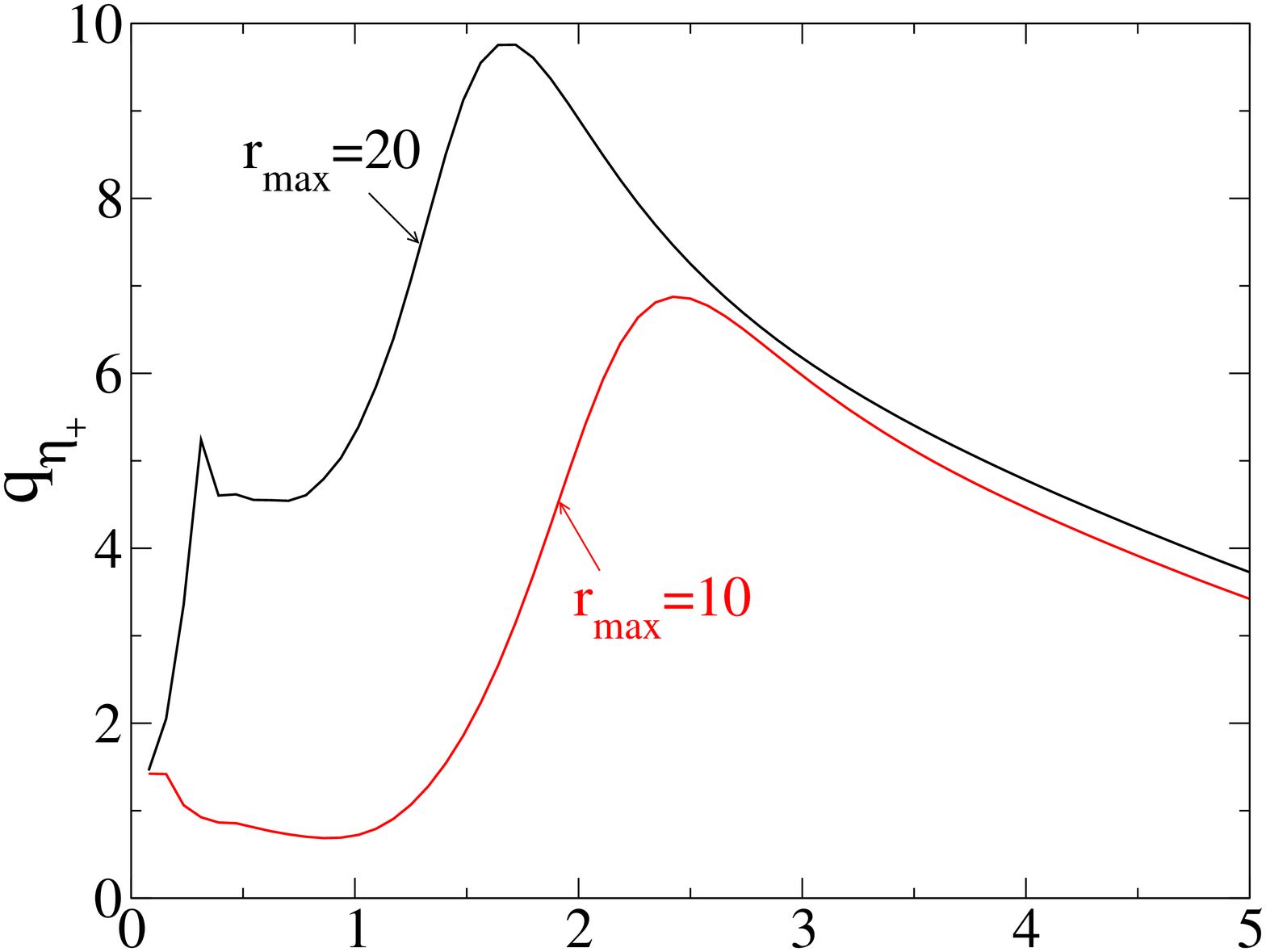}
    \medskip\\
    \includegraphics[width=0.95\textwidth]{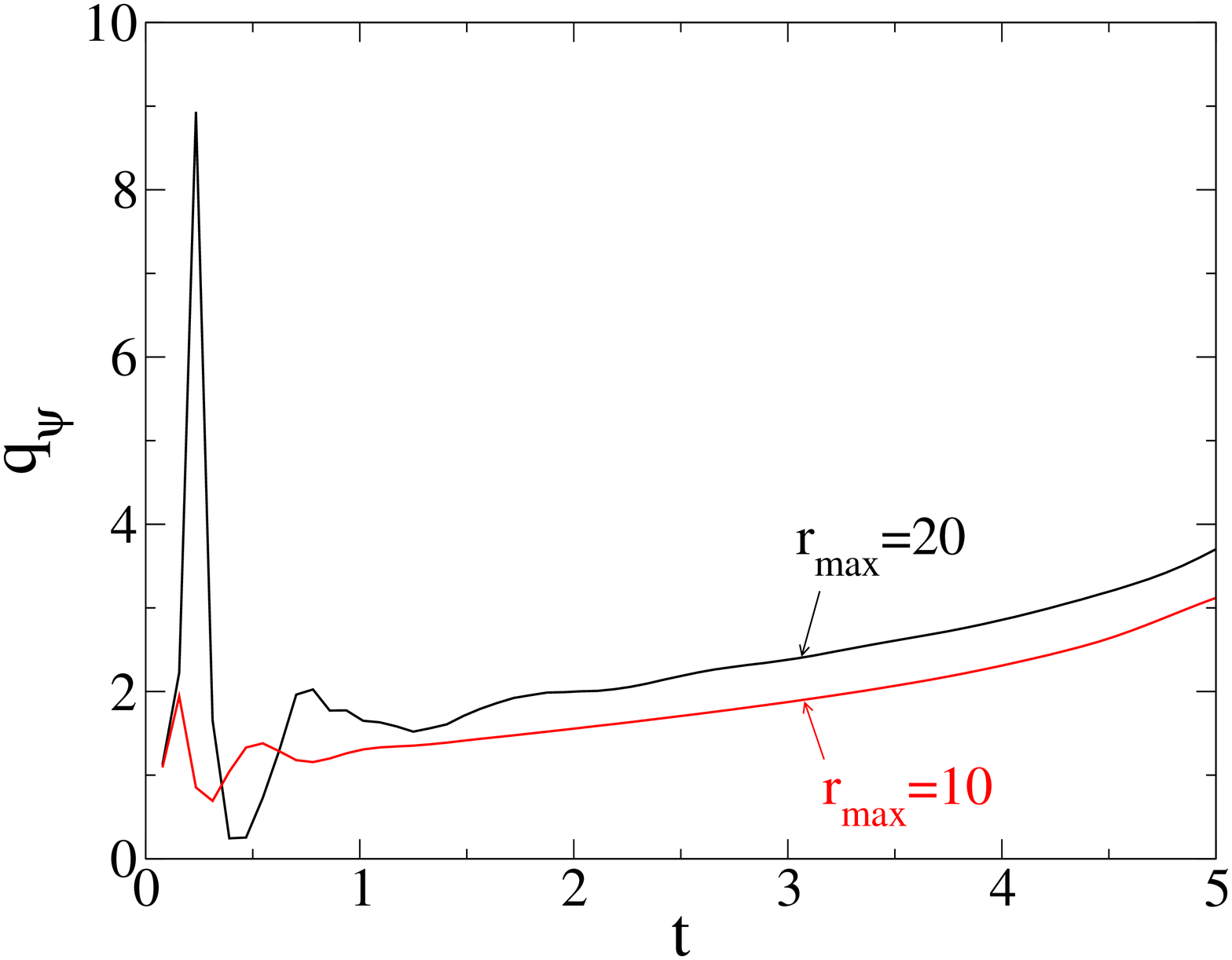}
  \end{minipage}
  \begin{minipage}[t]{0.5\textwidth}
    \includegraphics[width=0.95\textwidth]{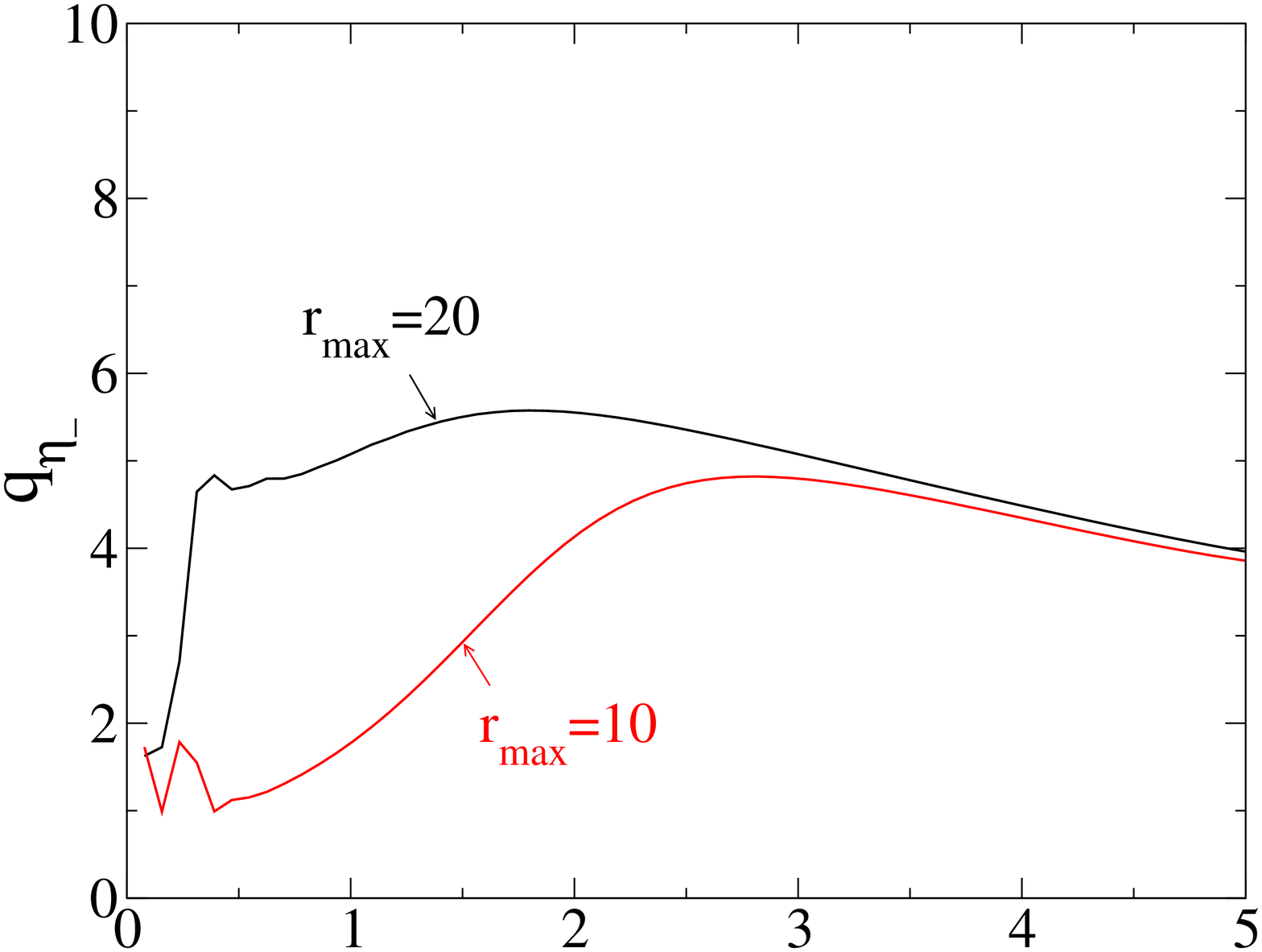}
    \medskip\\
    \includegraphics[width=0.95\textwidth]{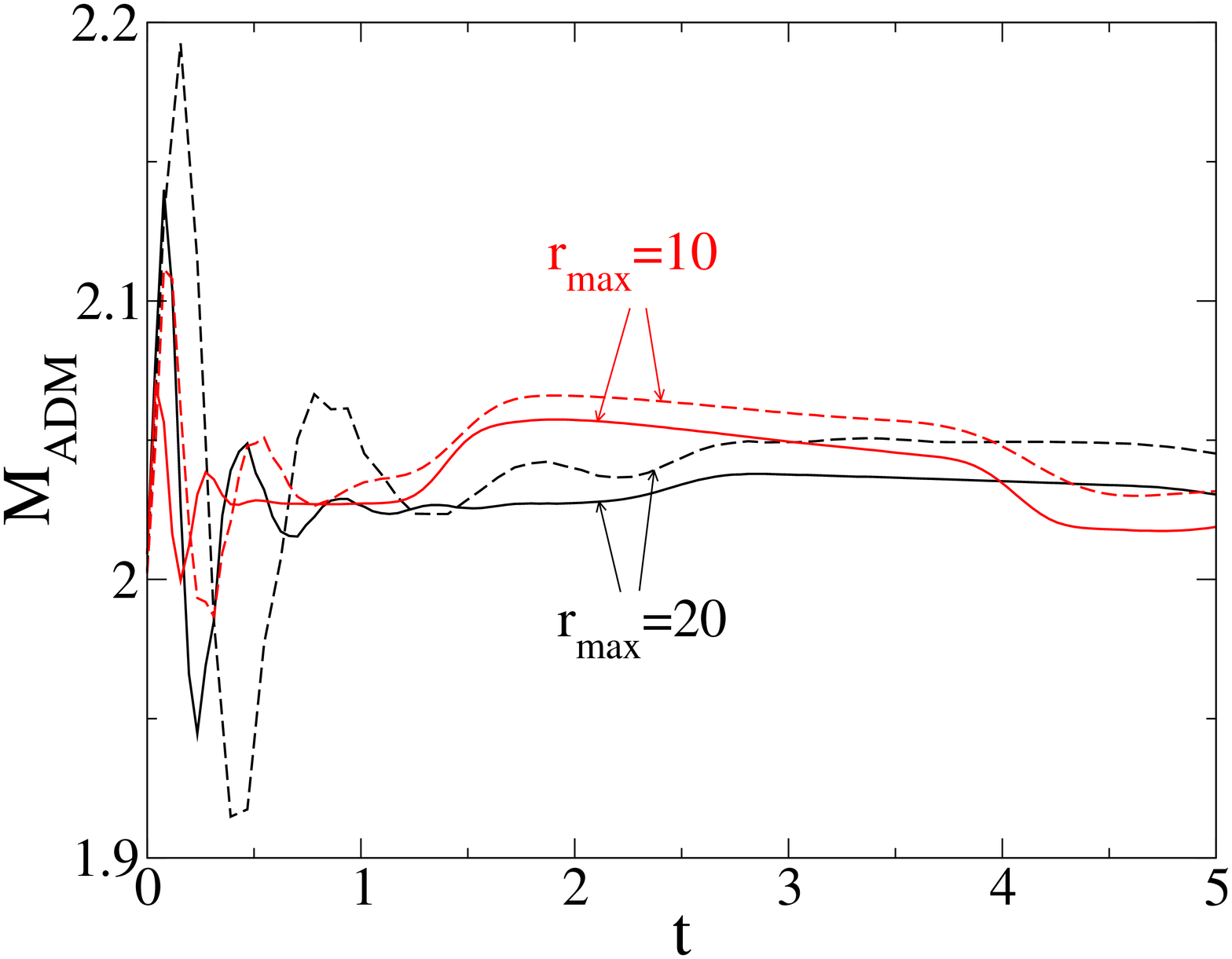}
  \end{minipage}
  \caption{\label{f:8.5resconv}
    Residual convergence factors \eref{e:q} for a Brill wave with 
    $\sigma_r = \sigma_z = 1$ and $a = 8.5$, for two different domain sizes 
    $r_\mathrm{max} = z_\mathrm{max}$. 
    The bottom right panel shows the numerically computed ADM mass 
    for the two different resolutions, $N_r = N_z = 128$ (dashed) 
    and $256$ (solid).
  }
\end{figure}

Figure \ref{f:8.5resconv} shows the residual convergence factors defined
in \eref{e:q}. 
The general trend is that the convergence factors are close to $4$ at late 
times but somewhat smaller at early times. 
Moving the outer boundary further out improves convergence considerably 
at early times, as can be seen particularly for the variables $\eta_\pm$. 
This demonstrates the effect of the outer boundary where imperfect
boundary conditions are imposed at a finite distance.
Because of the elliptic equations involved in our evolution scheme,
inaccuracies in the outer boundary conditions influence the entire 
domain instantaneously, not only after the outgoing radiation interacts 
with the boundary as is the case in a hyperbolic scheme.
Moving the outer boundary much further out by adding more coarse grids
is not feasible for this evolution because of the computational cost
involved in the current single-processor implementation of the code.

In particular, the value of the conformal factor $\psi$ far out
appears to be very sensitive to the dynamics close to the origin. 
This is not the case for $\hat \psi$, which is evolved from the
initial data by a hyperbolic PDE. 
As a result, the difference $\hat \psi - \psi$ has a large, spatially 
nearly constant contribution that is nearly resolution independent, 
thus causing the convergence factor $q_\psi$ to degrade.

At times later than those shown here, the convergence factors
ultimately decrease because large gradients develop due to the
grid-stretching property of maximal slicing. 
However, here we are only interested in the part of the evolution
until just after the formation of the apparent horizon.

Figure \ref{f:8.5resconv} also indicates that both increasing the 
resolution and the boundary radius improves the numerical conservation 
of the ADM mass. 
For the larger domain at the higher resolution, the initial oscillations 
are at the $5\%$ level and after $t \approx 1$ the mass remains constant 
to within $1\%$.

\begin{figure}
  \begin{minipage}[t]{0.5\textwidth}
    \includegraphics[width=0.95\textwidth]{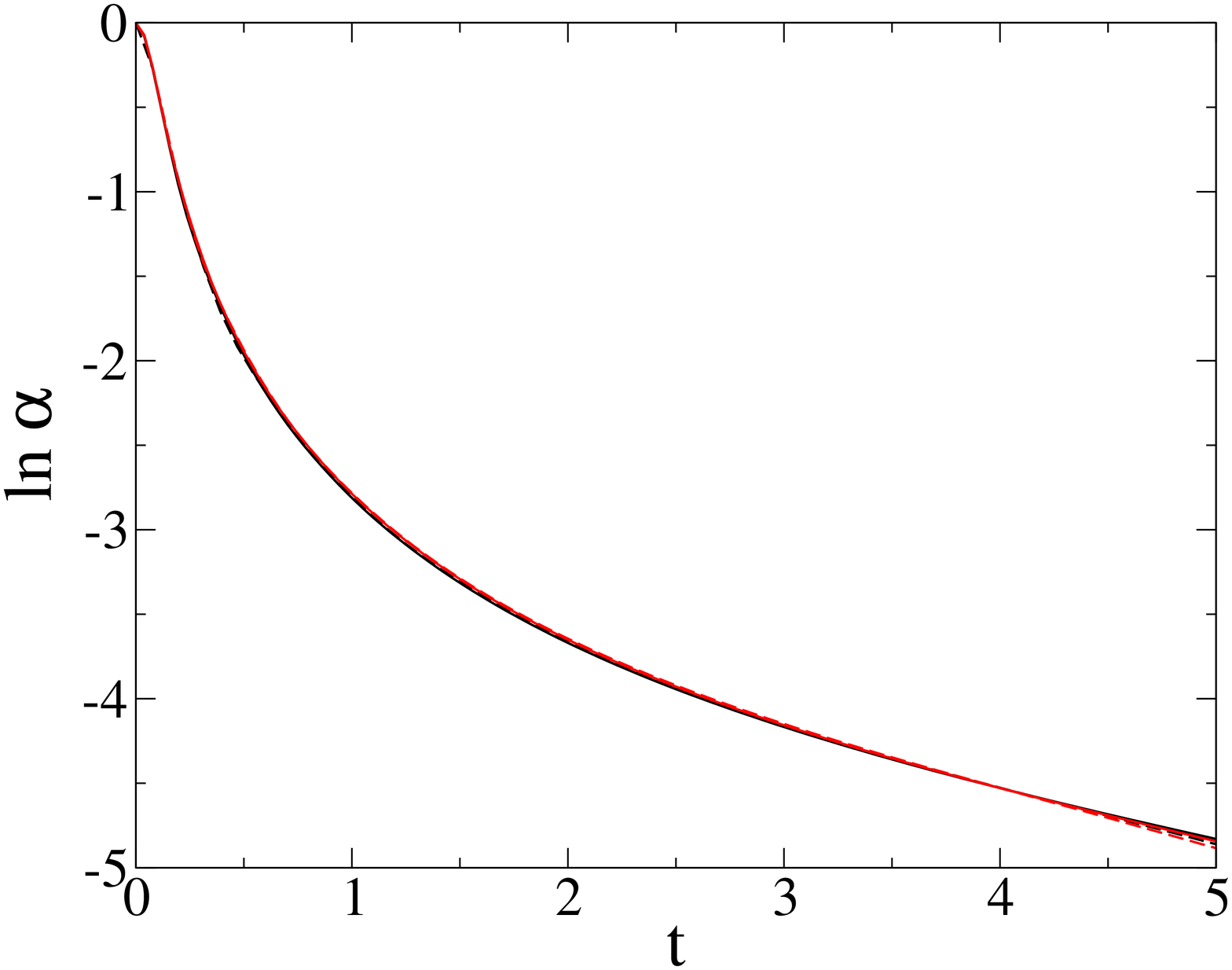}
  \end{minipage}
  \begin{minipage}[t]{0.5\textwidth}
    \includegraphics[width=0.95\textwidth]{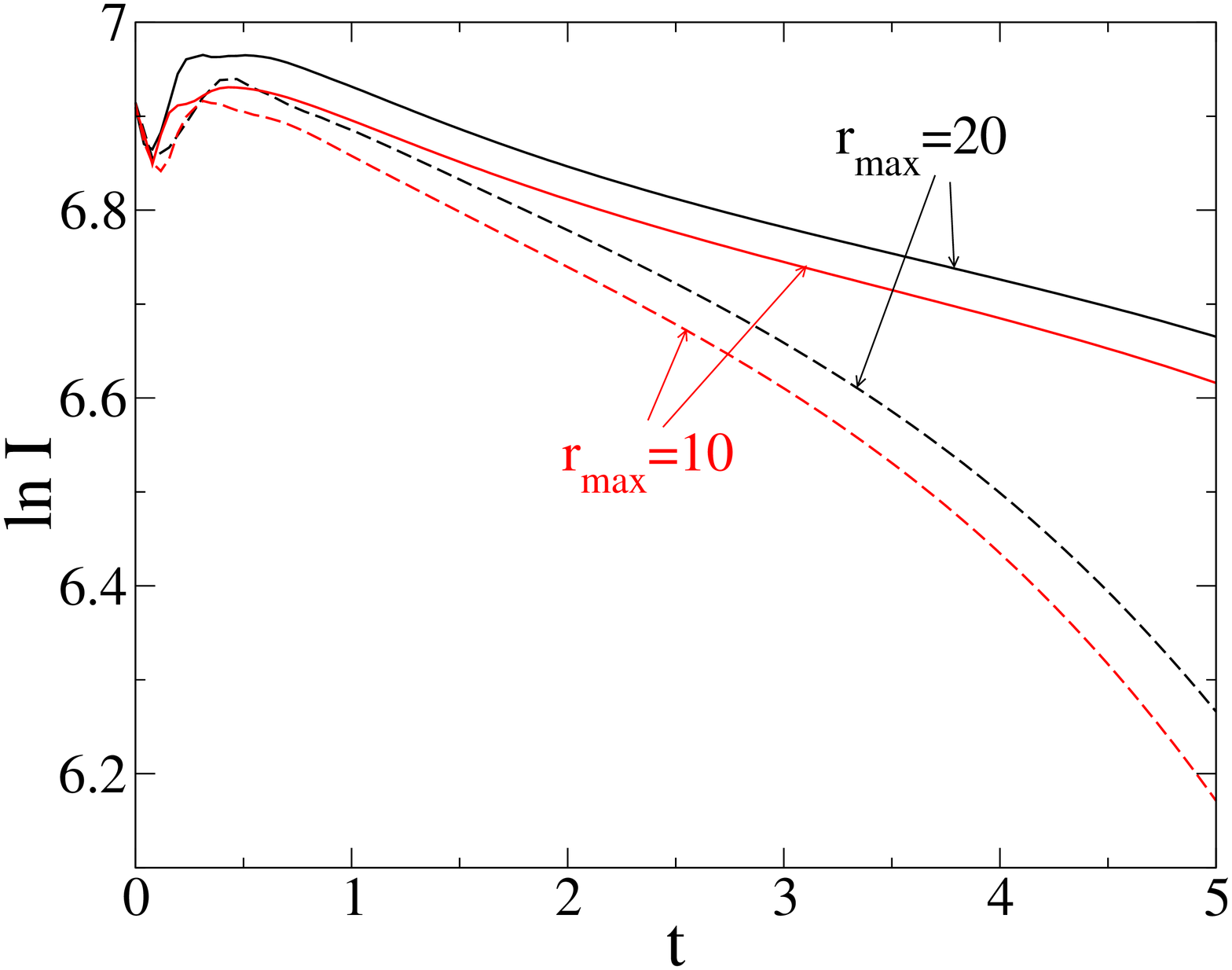}
  \end{minipage}
  \caption{\label{f:8.5origin}
    Lapse function $\alpha$ and Riemann invariant $I$ in the origin
    for a Brill wave with $\sigma_r = \sigma_z = 1$ and $a = 8.5$.
    Results for two different domain sizes $r_\mathrm{max} = z_\mathrm{max}$
    and for two different resolutions $N_r = N_z = 128$ (dashed) and
    $256$ (solid) are shown. (The four curves nearly coincide in the
    left plot.)
  }
\end{figure}

Next we evaluate the lapse function $\alpha$ in the origin $r = z = 0$. 
As a consequence of the singularity avoidance property of maximal 
slicing, the lapse is expected to collapse towards zero as a 
strong-gravity region of spacetime is approached. 
Our result in figure \eref{f:8.5origin} is in good agreement 
with \cite{Garfinkle2001} and appears to be insensitive to the
resolution and boundary location.

We also plot the Riemann invariant $I = R^{abcd} R_{abcd} / 16$
in the origin. The decay of this quantity after $t \approx 1$ agrees
roughly with \cite{Garfinkle2001}, although we find a somewhat different
behaviour at earlier times (rather than increasing right from the
beginning, $I$ first decreases for a short time). However there is a
rather strong dependence on resolution and outer boundary location
in this case, which indicates that the results for $I$ should be 
interpreted with care.

\begin{figure}
  \begin{minipage}[t]{0.5\textwidth}
    \includegraphics[width=0.95\textwidth]{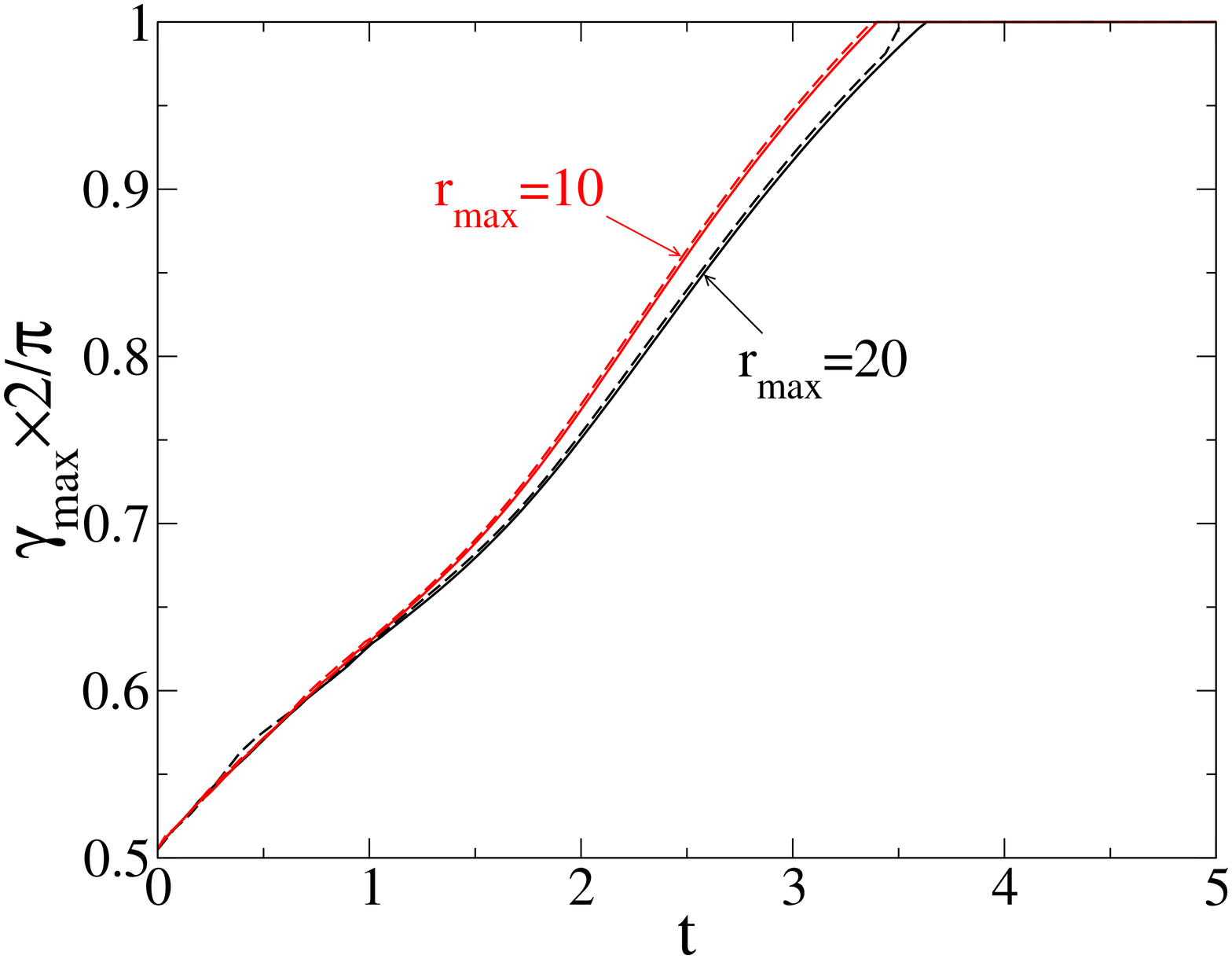}
  \end{minipage}
  \begin{minipage}[t]{0.5\textwidth}
    \includegraphics[width=0.95\textwidth]{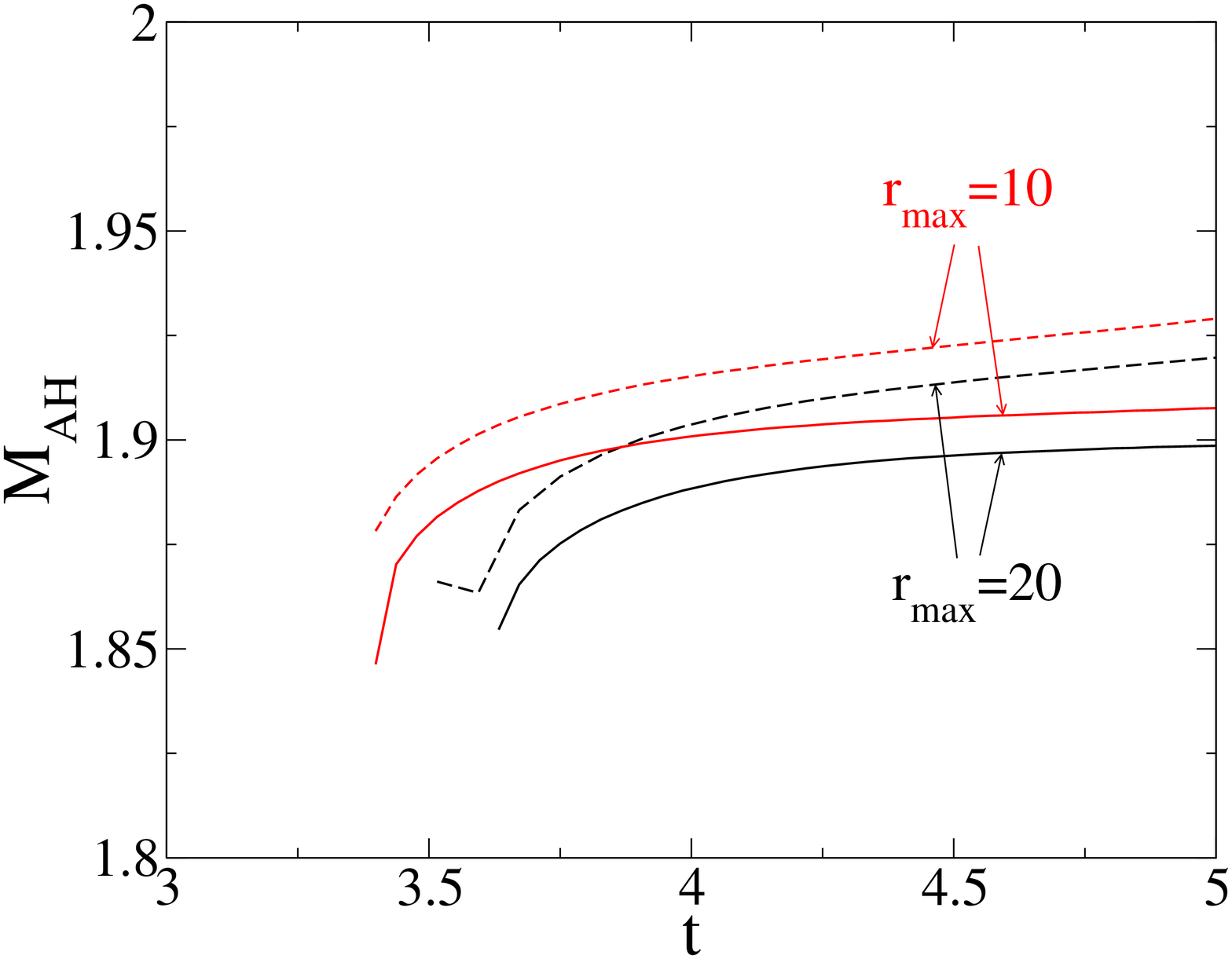}
  \end{minipage}
  \caption{\label{f:8.5horizon}
    Apparent horizon finder angle $\gamma_\mathrm{max}$ 
    (cf.~\eref{e:GammaAngle}) and mass $M_\mathrm{AH}$ for a Brill
    wave with $\sigma_r = \sigma_z = 1$ and $a = 8.5$.
    Results for two different domain sizes $r_\mathrm{max} = z_\mathrm{max}$
    and for two different resolutions $N_r = N_z = 128$ (dashed) and
    $256$ (solid) are shown.
  }
\end{figure}

Finally we search for an apparent horizon. The evolution of the angle 
$\gamma_\mathrm{max}$ (cf.~\eref{e:GammaAngle}) is shown in figure
\ref{f:8.5horizon}. It agrees reasonably well with \cite{Garfinkle2001},
although we find that the horizon forms slightly earlier at 
$t = 3.6 \pm 0.2$ rather than at $t = 3.9$. 
Also shown in figure \ref{f:8.5horizon} is the mass of the horizon, 
computed from its area $A_\mathrm{AH} = 16 \pi M_\mathrm{AH}^2$.
When it first forms, the horizon has mass $M_\mathrm{AH} = 1.85 \pm 0.05$.
The numerically computed ADM mass at this time is $M_\mathrm{ADM} =
2.04 \pm 0.02$ so that 
$M_\mathrm{AH}/M_\mathrm{ADM} = 0.91 \pm 0.03$, as compared with
$M_\mathrm{AH}/M_\mathrm{ADM} = 0.82$ reported in \cite{Garfinkle2001}.
After its formation the horizon expands slightly (its mass increases by about
$3\%$) and appears to ultimately settle down. 
The results stated here correspond to the run on the larger domain at
the higher resolution and the errors are estimated by comparison with
the other runs.

\subsection{Highly prolate collapse}
\label{s:ProlateWave}

We now turn to a highly prolate Brill wave with $\sigma_r = 0.128$, 
$\sigma_z = 1.6$, and $a = 325$, which again has $M_\mathrm{ADM}
\approx 2$. This is one of the initial data sets considered
in \cite{Abrahams1992} and it was also evolved (until $t \approx
1.5$) in \cite{Garfinkle2001}.

Our spatial domain has size $r_\mathrm{max} = z_\mathrm{max} = 20$.
The resolution on the base grid is taken to be $N_r = 256$ and $N_z = 64$.
There are $6$ grids in the FMR hierarchy.
Again all the grids contain the origin and the grid spacing is successively
halved in both dimensions. The number of grid points $N_r$ in the
radial direction is the same on all grids but $N_z$ is successively
multiplied by a factor of (approximately) $1.34$. In this way the
finer grids are better adapted to the prolate shape of the initial
data. The grid hierarchy is shown in figure \ref{f:FMRhierarchy}.
The spacing on the finest grid is $\Delta r = 2.44\times 10^{-3}$ and 
$\Delta z = 9.77\times 10^{-3}$. By comparison, the grid used 
in \cite{Garfinkle2001} has $\Delta r = 9.70\times 10^{-3}$ and 
$\Delta z = 3.74\times 10^{-2}$ close to the origin, roughly four
times coarser.

\begin{figure}
  \begin{minipage}[t]{0.5\textwidth}
    \includegraphics[width=0.95\textwidth]{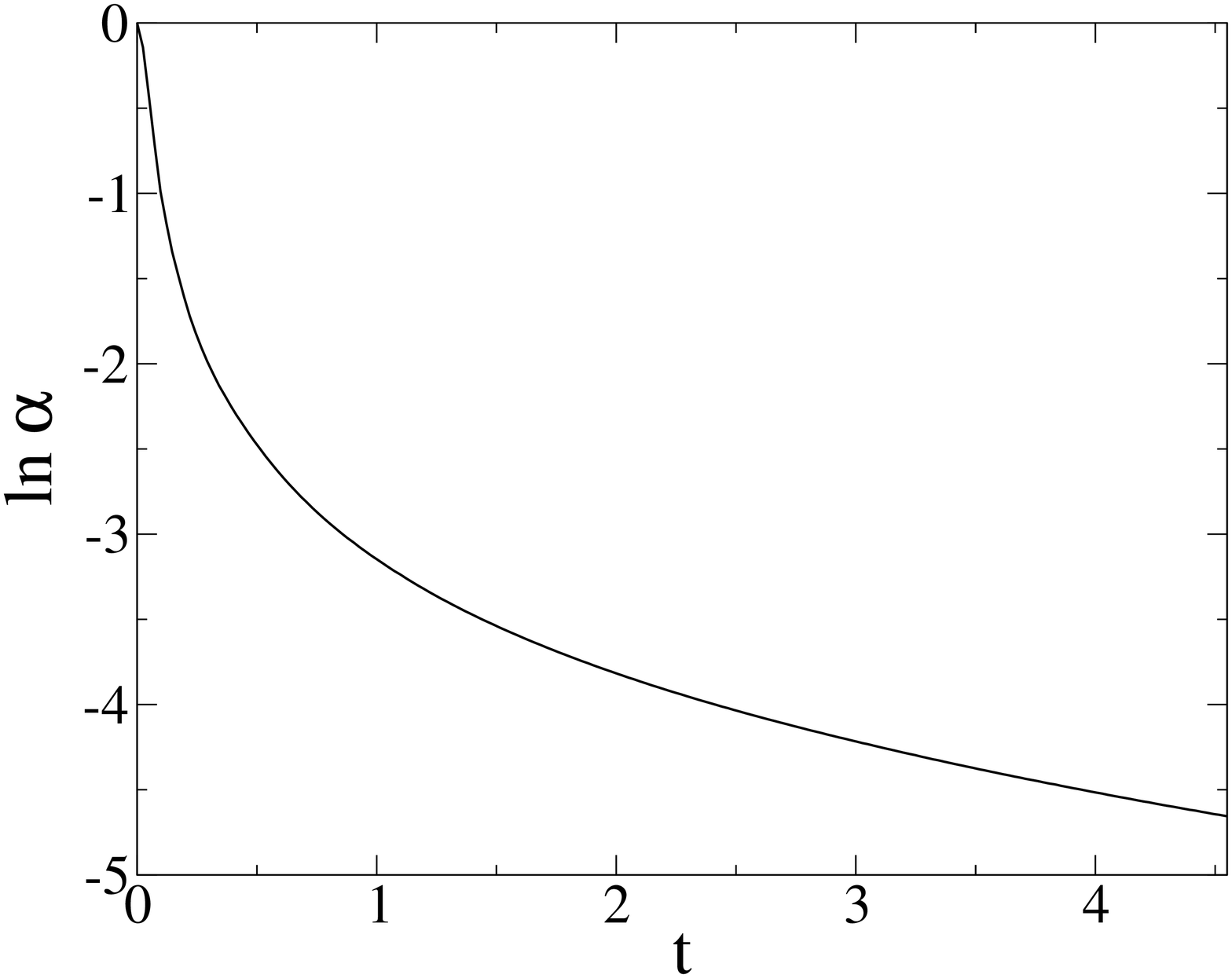}
  \end{minipage}
  \begin{minipage}[t]{0.5\textwidth}
    \includegraphics[width=0.95\textwidth]{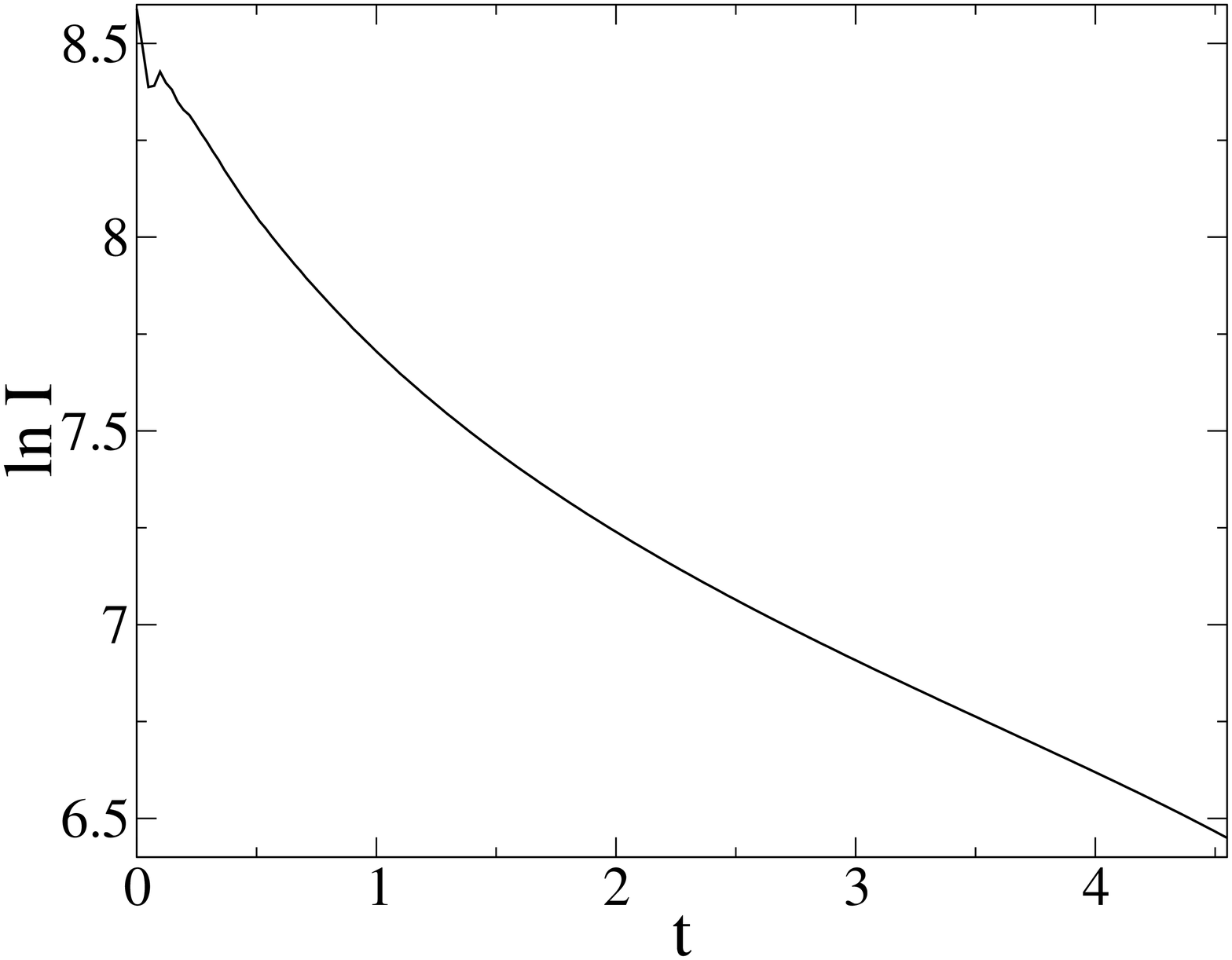}
  \end{minipage}
  \caption{\label{f:325origin}
    Lapse function $\alpha$ and Riemann invariant $I$ in the origin
    for a Brill wave with $\sigma_r = 0.128$, $\sigma_z = 1.6$ and $a = 325$.
  }
\end{figure}

Figure \ref{f:325origin} shows the evolution of the lapse function 
$\alpha$ and Riemann invariant $I$ in the origin. These agree well
with \cite{Garfinkle2001}, except for the $t \lesssim 0.5$ part of $I$
(but note the strong dependence of this quantity on resolution and 
outer boundary location apparent from figure \ref{f:8.5origin}).

\begin{figure}
  \begin{minipage}[t]{0.5\textwidth}
    \includegraphics[width=0.95\textwidth]{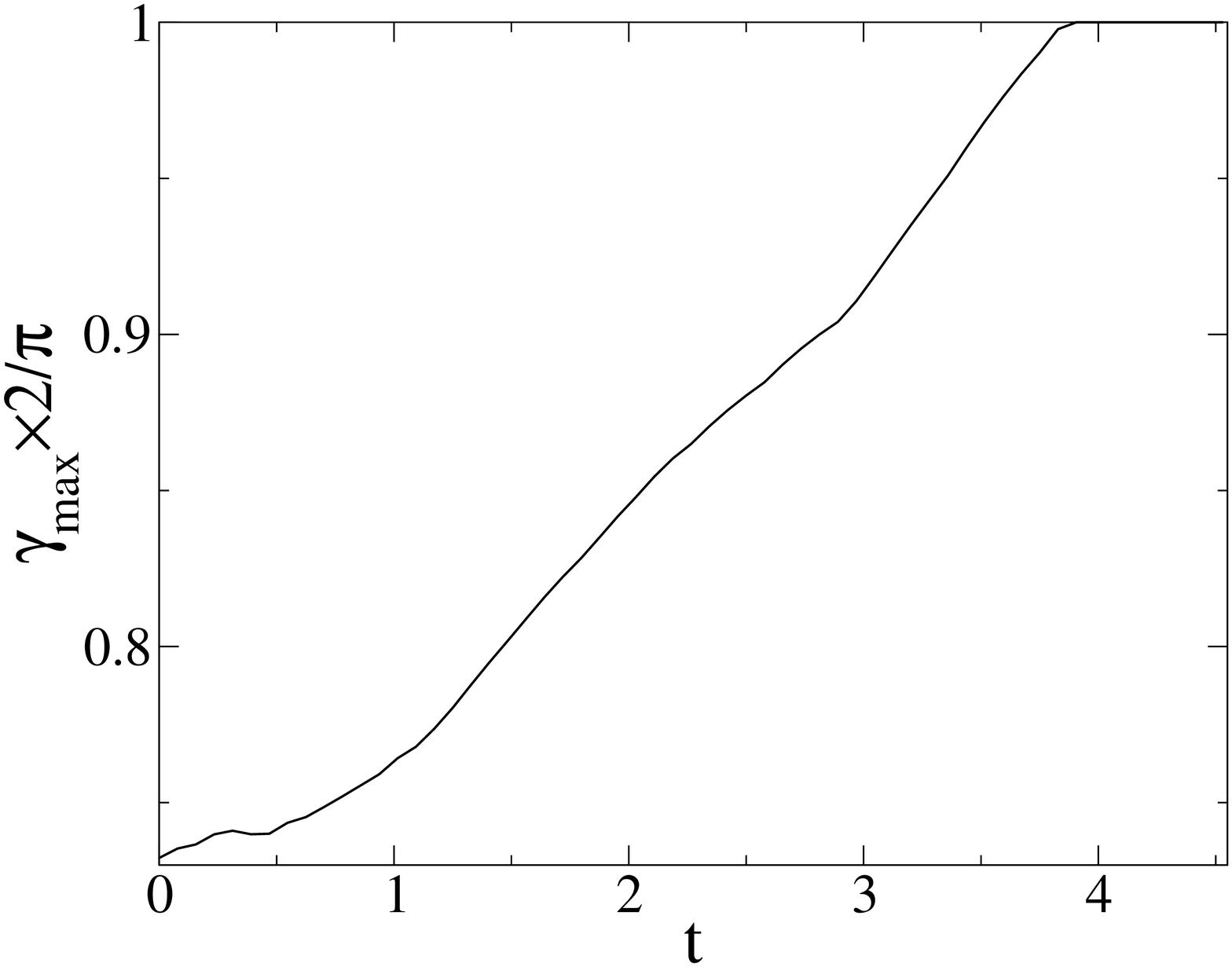}
  \end{minipage}
  \begin{minipage}[t]{0.5\textwidth}
    \includegraphics[width=0.95\textwidth]{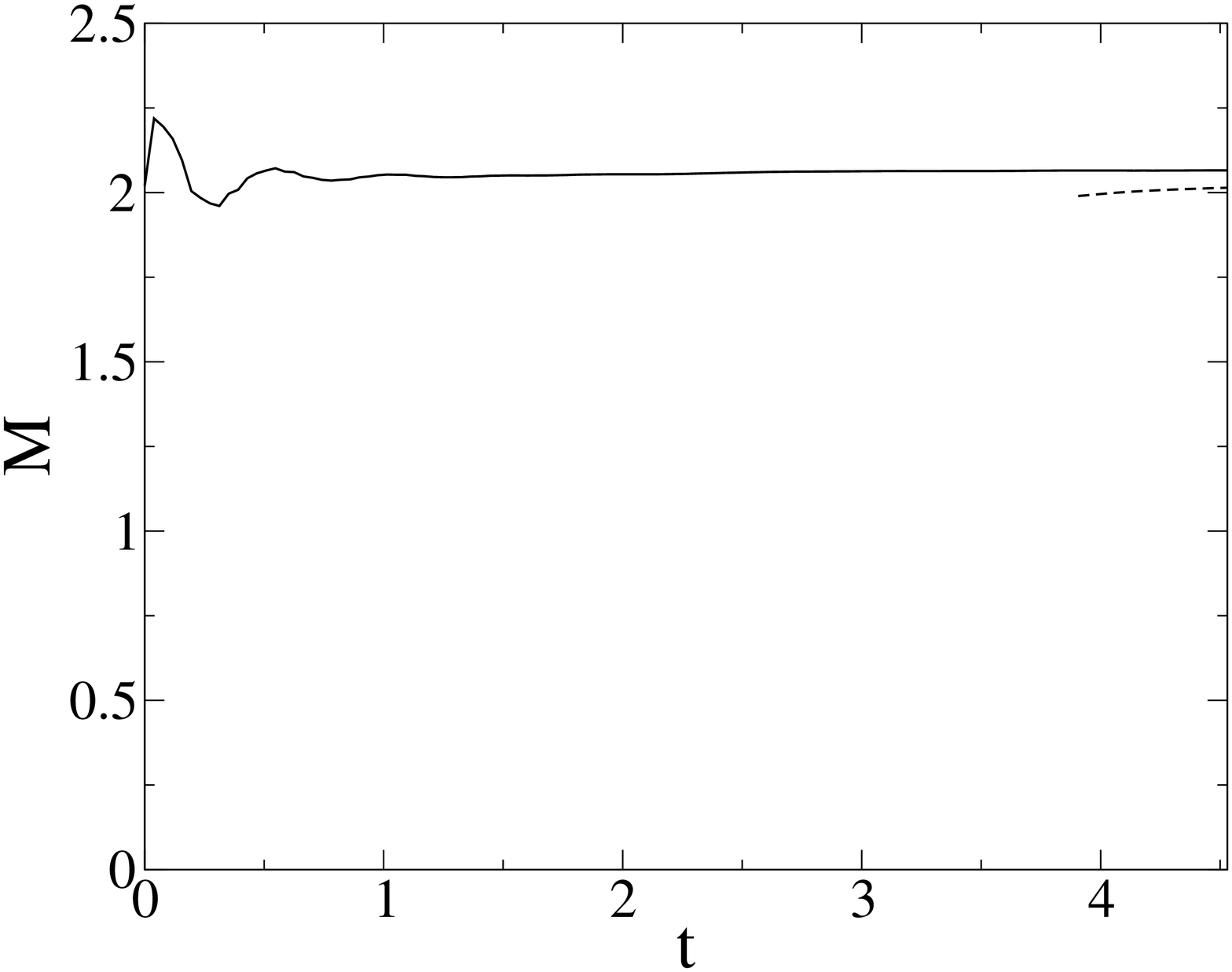}
    \vspace{0.8cm}\\
    \includegraphics[width=0.85\textwidth]{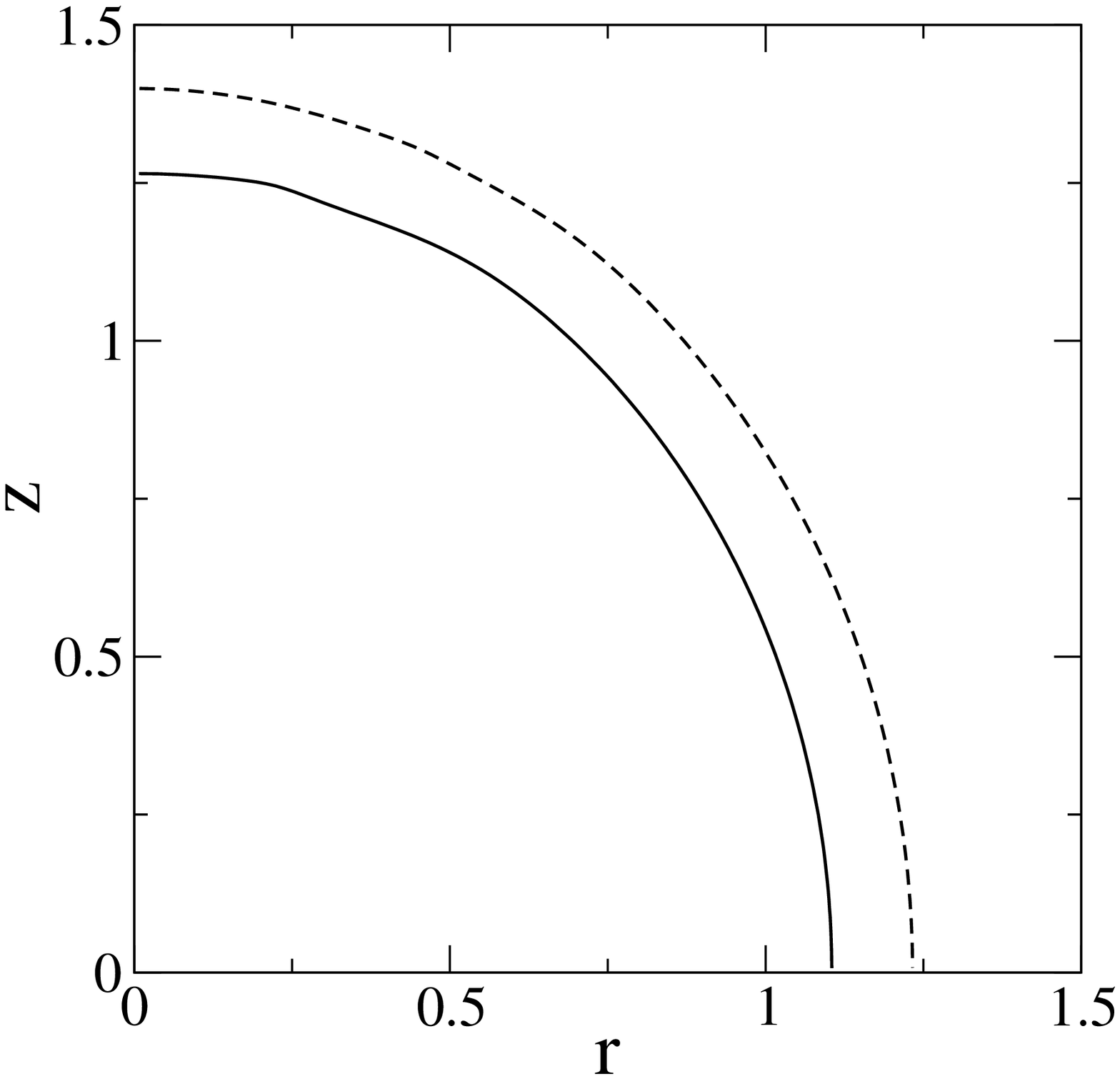}
  \end{minipage}
  \caption{\label{f:325horizon}
    Apparent horizon formation for a Brill wave with $\sigma_r = 0.128$, 
    $\sigma_z = 1.6$ and $a = 325$.
    Top left: horizon finder angle $\gamma_\mathrm{max}$ 
    (cf.~\eref{e:GammaAngle}).
    Top right: ADM mass (solid) and apparent horizon mass (dashed).
    Bottom: coordinate location of the apparent horizon when it first forms 
    at $t = 3.9$ (solid), and at $t = 4.5$ (dashed).
  }
\end{figure}

The approach to apparent horizon formation is shown in 
figure \ref{f:325horizon}.
We are able to evolve the wave for much longer than the authors 
of \cite{Garfinkle2001} and we confirm their conjecture that an
apparent horizon will indeed form.
It first appears at $t = 3.9$ and its shape is remarkably close to a
sphere in our coordinates.
At its formation the apparent horizon has mass $M_\mathrm{AH} = 1.990$ 
and at this time the ADM mass has settled down to a value of
$M_\mathrm{ADM} = 2.065$ so that $M_\mathrm{AH}/M_\mathrm{ADM} = 0.96$.
This is in accordance with the Penrose inequality \cite{Penrose1973}, 
which conjectures that $M_\mathrm{AH}/M_\mathrm{ADM} \leqslant 1$.

%%%%%%%%%%%%%%%%%%%%%%%%%%%%%%%%%%%%%%%%%%%%%%%%%%%%%%%%%%%%%%%%%%%%%%%%%%%%%%%

\section{Conclusions}
\label{s:Concl}

We considered constrained evolution schemes for the Einstein equations
in axisymmetric vacuum spacetimes.
One of the motivations for this work was to try and understand why the
numerical elliptic solvers in some of these schemes, 
e.g.~\cite{Choptuik2003}, failed to converge in certain situations. 
We found that this was related to the elliptic equations becoming
indefinite.
Apart from the implications for numerical convergence, we also pointed out
that such equations might admit nonunique solutions.
In section \ref{s:ORscheme}, we presented a new scheme that does not
suffer from this problem. 
Its main features are a suitable rescaling of the extrinsic curvature
with the conformal factor, and separate solution of the momentum
constraints and isotropic spatial gauge conditions.
Thus the scheme involves the solution of six elliptic equations rather 
than four as in \cite{Choptuik2003}.
Given that multigrid methods \cite{Brandt1977} can be used to solve
these equations at linear complexity, this does not imply a severe increase
in computational cost.

Our numerical implementation uses second-order accurate finite
differences and combines mesh refinement with a multigrid elliptic
solver, based on the algorithm of \cite{Pretorius2006a}.
We work in cylindrical polar coordinates.
Unlike in \cite{Garfinkle2001}, we do not compactify the spatial
domain but impose boundary conditions at a finite distance from the origin.

As an application of the code, we evolved Brill waves in section 
\ref{s:NumResults}. We carried out a careful convergence test in 
section \ref{s:ConvTest} and demonstrated that the code is 
approximately second-order convergent.
For a stronger super-critical wave (section \ref{s:SpherWave}), 
convergence of the residuals of the unsolved evolution equations was 
somewhat slower at earlier times. 
Varying the domain size indicated that this is mainly caused by 
inaccuracies in the outer boundary conditions we use. 
These errors appear to have little effect on the formation of the
apparent horizon.

In section \ref{s:ProlateWave}, we evolved a highly prolate Brill wave.
Such initial data were conjectured in \cite{Abrahams1992} to form a
naked singularity rather than a black hole, which would violate weak
cosmic censorship.
However an apparent horizon does form in our evolution.
We thus improve on the results of the authors of \cite{Garfinkle2001},
who could not evolve the wave for a sufficiently long time to see an
apparent horizon form although they conjectured that this would happen
eventually.

There are many directions in which this work could be extended.
For simplicity we only considered vacuum spacetimes with a
hypersurface-orthogonal Killing vector, i.e., vanishing \emph{twist}. 
The addition of matter and twist should be straightforward.
Care must be taken that the additional variables are rescaled with
suitable powers of the conformal factor so that the Hamiltonian
constraint \eref{e:ORHamiltonianConstraint} remains definite.
An elegant framework capable of including twist is provided by the $(2+1)+1$ 
formalism \cite{Geroch1971,Maeda1980}.
From a mathematical point of view, it would be interesting to prove
that the Cauchy problem or even the initial-boundary value problem is
well posed for the present (or a similar) formulation of the
axisymmetric Einstein equations.
These questions were studied for similar hyperbolic-elliptic
systems in \cite{ChoquetBruhat1996,Andersson2003,Nagy2006}.

It is a disadvantage of constrained evolution schemes that
inaccuracies in the outer boundary conditions influence the entire 
domain instantaneously.
More work is needed on improved boundary conditions in the context of
mixed hyperbolic-elliptic formulations of the Einstein equations.
An alternative to an outer boundary at a finite distance would be the
compactification towards spatial infinity used in \cite{Garfinkle2001}.
However, outgoing waves ultimately fail to be resolved on such a
compactified grid, which because of the elliptic equations involved
has again adverse effects on the entire solution.
This problem is avoided if hyperboloidal slices are used, which
can be compactified towards future null infinity (see \cite{FrauendienerLRR}
for a related review article). In this case the constraint and
evolution equations become formally singular at the boundary, 
which needs to be addressed in a numerical implementation.

On the computational side, the accuracy of the code could be improved by 
using fourth (or higher) order finite differences. Computational speed could
be gained by parallelizing the code and running it on multiple processors.

It would be interesting to evolve even more prolate Brill waves than
the one considered here.
However the elliptic equations then become more and more anisotropic
and the relaxation method employed in the multigrid method must be
modified to ensure convergence. 
For the wave considered in this paper, line relaxation was found to
accomplish this but we have not been able to achieve convergence for
even more prolate configurations. 
More sophisticated modifications such as operator-based prolongation and
restriction \cite{Trottenberg2001} are likely to be required.
In any case, in order to evolve some of the extremely prolate initial
data sets considered in \cite{Abrahams1992} where 
$\sigma_r/\sigma_z \approx 10^4$ in \eref{e:InitialData},
a radically different approach will probably be needed.

Another interesting application of our code would be Brill wave
critical collapse.
However, preliminary results indicate that we are currently unable to evolve
waves close to the critical point for a sufficiently long time because
reflections from the interior AMR grid boundaries become increasingly
pronounced as more and more finer grids are added close to the origin.
The mesh refinement algorithm of \cite{Pretorius2006a} that we adopt
appeared to be sufficiently robust in the scalar field evolutions 
of \cite{Choptuik2003b,Choptuik2004} but we suspect that the situation
is quite different in vacuum collapse. 
An improved AMR algorithm for mixed hyperbolic-elliptic systems of
PDEs will probably be required.

%%%%%%%%%%%%%%%%%%%%%%%%%%%%%%%%%%%%%%%%%%%%%%%%%%%%%%%%%%%%%%%%%%%%%%%%%%%%%%%

\ack 
The author wishes to thank Sergio Dain, David Garfinkle, John Stewart 
and Darragh Walsh for helpful discussions on this work.
He gratefully acknowledges funding through a Research Fellowship at
King's College Cambridge.
Earlier parts of this work were supported by grants to Caltech from the
Sherman Fairchild Foundation, NSF grant PHY-0601459 and NASA grant
NNG05GG52G. 

%%%%%%%%%%%%%%%%%%%%%%%%%%%%%%%%%%%%%%%%%%%%%%%%%%%%%%%%%%%%%%%%%%%%%%%%%%%%%%%

\appendix

\section{Evolution equations}
\label{s:EvolutionEquations}

Here we list the evolution equations of the formulation of the
axisymmetric Einstein equations presented in section \ref{s:ORscheme}.
The variables $S$ and $\tW$ are actively evolved,
\begin{eqnarray}
  \label{e:EvolutionS}
  \fl S_{,t} = \beta^r S_{,r} + \beta^z S_{,z} - \alpha \psi^{-6} \tW 
    + r^{-1} \beta^r S + (r^{-1} \beta^r)_{,r} \; , \\
  \label{e:EvolutionW}
  \fl \tW_{,t} = \beta^r \tW_{,r} + \beta^z \tW_{,z} + r^{-1} \eta_+
    \beta^z{}_{,r}  
    - \alpha \psi^2 \rme^{-2rS} \left[ \alpha^{-1} (r^{-1} \alpha_{,r})_{,r} 
      + 2 \psi^{-1} (r^{-1} \psi_{,r})_{,r} \right.\nonumber\\\left.
      + S_{,rr} + S_{,zz} + (r^{-1} S)_{,r}
      - r^{-1} R_r (A_r + 2 P_r) + r^{-1} R_z (A_z + 2 P_z) \right.\nonumber\\\left.
      - 2 r^{-1} P_r (2 A_r + 3 P_r) \right] 
    \nonumber\\
    - \half \alpha \psi^{-6} r^{-1} [\eta_+^2 + 2 \eta_- r\tW - 4 (r\tW)^2 ]
      + 4 r^{-1} \beta^r \tW.
\end{eqnarray}
The variables $\psi$ and $\eta^A$ are not evolved but solved for
using the constraint equations. However the Einstein equations also
imply the following evolution equations that may be used to check
the accuracy of a numerical scheme.
\begin{eqnarray}
  \label{e:psievolution}
  \fl \psi_{,t} = \beta^r \psi_{,r} + \beta^z \psi_{,z} + \half r^{-1}
    \beta^r \psi - \tfrac{1}{6} \alpha \psi^{-5} (\eta_- - 2 r \tW), \\
  \label{e:etapevolution}  
  \fl \eta_{+,t} = 2 \alpha \psi^2 \rme^{-2rS} [ -\alpha^{-1} \alpha_{,rz} 
    - 2 \psi^{-1} \psi_{,rz} + R_r A_z + R_z (A_r + r^{-1}) + 6 P_r P_z 
    \nonumber\\ 
    + 2 P_r (A_z + R_z) + 2 P_z (A_r + R_r) ] \nonumber\\
    + \beta^r (\eta^r{}_{,rz} + \eta^z{}_{,rr} + 3 r^{-1} \eta_+)
    + \beta^z (\eta^r{}_{,zz} + \eta^z{}_{,rz}) - \eta_+ \beta_- \nonumber\\
    + \tfrac{2}{3} (\beta^r{}_{,z} - 2 \beta^z{}_{,r}) \eta_-
    + \tfrac{2}{3} r \tW \beta_+ 
    + \third \alpha \psi^{-6} \eta_+ (\eta_- + 4 r \tW) , \\
  \label{e:etamevolution}  
  \fl \eta_{-,t} = \alpha \psi^2 \rme^{-2rS} [ 
    - \alpha^{-1} \alpha_{,rr} + \alpha^{-1} \alpha_{,zz} 
    - 2 \psi^{-1} \psi_{,rr} + 2 \psi^{-1} \psi_{,zz} + 2 R_r (A_r + r^{-1}) 
    \nonumber\\ 
    - 2 A_z R_z + 2 P_r (2 A_r + 3 P_r + 2 R_r)  
    - 2 P_z (2 A_z + 3 P_z + 2 R_z) ] \nonumber\\
    + \beta^r (\eta^r{}_{,rr} - \eta^z{}_{,rz} + 3 r^{-1} \eta_-)
    + \beta^z (\eta^r{}_{,rz} - \eta^z{}_{,zz}) \nonumber\\
    - \tfrac{2}{3} \beta_- (2 \eta_- - r \tW) 
    + (\beta^z{}_{,r} - \beta^r{}_{,z}) \eta_+ 
    + \third \alpha \psi^{-6} \eta_- (\eta_- + 4 r \tW) .
\end{eqnarray}

%%%%%%%%%%%%%%%%%%%%%%%%%%%%%%%%%%%%%%%%%%%%%%%%%%%%%%%%%%%%%%%%%%%%%%%%%%%%%%%

\section{Finite-difference operators and boundary conditions}
\label{s:Discretization}

Centred second-order accurate finite-difference operators are used at
all interior grid points.
We only give the expressions for derivatives in the $r$ direction;
corresponding expressions hold in the $z$ direction.
The symbol $\approx$ denotes equality up to $\Or(\Delta r)^2$.
\begin{eqnarray} 
  \label{e:FDoperators}
  u_{,r} \vert_i &\approx& (u_{i+1} - u_{i-1}) / (2 \Delta r),\nonumber\\
  (r^{-1} u)_{,r} \vert_i &\approx& 
    (u_{i+1}/r_{i+1} - u_{i-1}/r_{i-1}) / (2 \Delta r),\nonumber\\
  u_{,rr} \vert_i &\approx& (u_{i+1} - 2 u_i + u_{i-1}) / (\Delta r)^2,\nonumber\\
  (r^{-1} u_{,r})_{,r} \vert_i &\approx&
    [(u_{i+2} - u_i)/r_{i+1} - (u_i - u_{i-2})/r_{i-1}] / [ 4 (\Delta r)^2 ].
\end{eqnarray}

At $r = 0$ ($z = 0$), a Dirichlet condition is imposed if the variable
is an odd function of $r$ ($z$) and a Neumann condition if it is an
even function of $r$ ($z$). The parities of the various variables are
as follows,
\begin{eqnarray}
  \textrm{odd in } r: & \beta^r, \eta^r, S, \tW \nonumber\\
  \textrm{even in } r: & \alpha, \psi, \beta^z, \eta^z \nonumber\\
  \textrm{odd in } z: & \beta^z, \eta^z \nonumber\\
  \textrm{even in } z: & \alpha, \psi, \beta^r, \eta^r, S, \tW  
\end{eqnarray}
The value of a variable $u$ at the ghost point is set to be 
$u_0 = - u_1$ if $u$ obeys a Dirichlet condition and 
$u_0 = u_1$ if it obeys a Neumann 
condition\footnote{The last operator in \eref{e:FDoperators} 
requires a ghost value $u_{-1}$; this is set according to 
$u_{-1} = \pm u_2$.}.
This discretization at the boundary is second-order accurate.

Dirichlet conditions at the outer boundary are implemented in a
similar way. In order to impose the falloff condition 
\eref{e:FalloffBC}, we note that $R u$ is a linear
function of $R$ and so we linearly extrapolate $R u$ in the radial
direction from the interior grid points to the ghost points in order 
to find the values of $u$ there. 
The Sommerfeld condition \eref{e:SommerfeldBC} is rewritten in the form
\begin{equation}
  R u_{,t} + r u_{,r} + z u_{,z} + u - u_\infty = 0
\end{equation}
and discretized at the outer ghost points on the base grid of the AMR 
hierarchy in order to integrate the values of $u$ there forward in time. 
Here backward differencing is used in the direction normal to the boundary,
\begin{equation}
  u_{,r} \vert_i \approx (3 u_i - 4 u_{i-1} + u_{i-2}) / (2 \Delta r) 
\end{equation}
and similarly in the $z$ direction.

%%%%%%%%%%%%%%%%%%%%%%%%%%%%%%%%%%%%%%%%%%%%%%%%%%%%%%%%%%%%%%%%%%%%%%%%%%%%%%%

\section*{References}

\providecommand{\newblock}{}

\end{document}